\documentclass[letterpaper,journal]{IEEEtran}
\usepackage{amsmath,amsfonts}
\usepackage{algorithmic}
\usepackage{algorithm}
\usepackage{array}
\usepackage[caption=false,font=normalsize,labelfont=sf,textfont=sf]{subfig}
\usepackage{textcomp}
\usepackage{stfloats}
\usepackage{url}
\usepackage{verbatim}
\usepackage{graphicx}
\usepackage{cite}
\newtheorem{definition}{\bf Definition}[section]

\newtheorem{remark}{\bf Remark}[section]

\newtheorem{proposition}{\bf Proposition}[section]

\usepackage{amssymb}
\usepackage{amsmath}
\usepackage{amsfonts} 
\usepackage{booktabs}
\usepackage{color}
\usepackage{enumerate}
\usepackage{tabularx}
\usepackage{makecell}
\usepackage{multirow}
\usepackage{indentfirst}
\usepackage{epsfig}

\begin{document}


\title{Online Monitoring and Risk Assessment of Non-Cooperative UAVs via STL-Aware Adaptive Fusion Kalman Filtering}

\author{
Xinhao~Yan,~\IEEEmembership{Graduate~Student~Member,~IEEE},
Ruige~Yang,
Chao~Peng,
Hailong~Huang,~\IEEEmembership{Senior~Member,~IEEE}
\thanks{This work was supported by the Multi-Sensory Data Fusion Approach and Integrated Physical Design of ISAC Base Stations, China Mobile Hong Kong Company Limited.}
\thanks{Xinhao~Yan, Ruige~Yang, and Hailong~Huang are with the Department of Aeronautical and Aviation Engineering, The Hong Kong Polytechnic University, Kowloon, Hong Kong (emails: xin-hao-shawn.yan@connect.polyu.hk; rrruige.yang@connect.polyu.hk; hailong.huang@polyu.edu.hk).}
\thanks{Chao~Peng is with DICT Centre, China Mobile Hong Kong Company Limited, 999077, Hong Kong, China (charlespeng@hk.chinamobile.com).}
\thanks{\emph{Corresponding author: Hailong Huang.}}
}

\markboth{IEEE Transactions on xx xx, vol. XX, no. XX, XX 2026}%
{Yan \MakeLowercase{\textit{et al.}}: Online Monitoring and Risk Assessment of Non-Cooperative UAVs via STL-Aware Adaptive Fusion Kalman Filtering}


\maketitle

\begin{abstract}
This paper considers the problem of online state estimation and predictive risk assessment for non-cooperative unmanned aerial vehicles (UAVs) in the presence of asynchronous heterogeneous sensing and uncertain motion modes.
To address this problem, a unified estimation and safety-assessment framework is developed by integrating an interacting multiple-model multi-rate Kalman filter with signal temporal logic (STL).
The proposed framework enables simultaneous low-level state tracking and high-level safety reasoning within a common recursive architecture.
Its main contribution is an STL-aware time-varying mode transition mechanism that updates model probabilities online using robustness measures induced by formal safety specifications.
By embedding safety semantics directly into the mode inference and estimation process, the method improves responsiveness to maneuver variations, sensing asynchrony, and evolving threat patterns.
Based on the estimated state distributions, the framework further generates multi-step state predictions and probabilistic reachable sets, which are used for finite-horizon safety evaluation and risk-triggered warning generation.
Consequently, the proposed method provides not only estimates of the current target state, but also early indication of unsafe behaviors before they become fully observable.
Finally, experimental results obtained from a real-time UAV monitoring platform show that the proposed approach improves estimation accuracy and produces earlier and more informative safety warnings, demonstrating its effectiveness for real-time UAV surveillance and safety monitoring applications.
\end{abstract}


\section{Introduction}

Owing to their excellent mobility and operational flexibility, unmanned aerial vehicles (UAVs) have become a powerful tool in many fields, including but not limited to modern farms \cite{Seo_Farms}, soil sampling \cite{Rao_Soil}, and intelligent transportation \cite{Antal_Transportation}.
This widespread adoption is driven by advancements in automation, sensing technologies, and cost reduction \cite{Arribas_Energy,Jin_Energy}, which enable UAVs to perform tasks that are dangerous, repetitive, and inaccessible to humans \cite{Wen_Dangerous}.
The rapid proliferation of UAVs, particularly non-cooperative ones that operate without prior authorization or adherence to airspace regulations \cite{Wan_Non}, poses escalating risks to public safety, privacy, and critical infrastructure \cite{XinhaoYan_Privacy_TASE,IchiroHasuo_Safety}.
Such UAVs may either intentionally or unintentionally enter restricted airspaces.
In this case, monitoring and warning systems are required to be capable of performing reliable trajectory tracking, multi-sensor data fusion, and real-time risk assessment with respect to predefined no-fly zones and operational constraints \cite{Ding_UAV,He_Non,Song_Non}.

Accurate state estimation of the UAV is essential before conducting risk assessment.
Numerous UAV monitoring systems rely on Kalman filtering techniques \cite{Zhu_IMMUKF,XinhaoYan_Aggregation_TCST} to fuse sensory data from sensors such as visual cameras \cite{Abdelkader_KF,Wang_Camera,Li_Camera}, radars \cite{Pardhasaradhi_KF,Ye_KF}, and infrared equipment \cite{Ye_Infrared}.
For instance, in one typical setup, four radars are deployed at the edge of a region to capture UAV trajectories \cite{Pardhasaradhi_KF}. 
However, practical monitoring scenarios often involve multi-rate and asynchronous measurements from heterogeneous sensors with unknown cross-correlations.
This poses a challenge for designing scalable and adaptive fusion frameworks that go beyond single-sensor cases \cite{Abdelkader_KF} or single-sensor-type configurations \cite{XinhaoYan_Aggregation_TCST,Pardhasaradhi_KF}.
In parallel, conventional accurate estimation methods generally require the full knowledge of the dynamics \cite{XinhaoYan_Estimation_TAES} and control inputs \cite{XinhaoYan_Estimation_TIV}.
Since these internal data of a non-cooperative UAV are generally unknown, accurate prediction based on the state transition model becomes infeasible.
Consequently, it is necessary to make assumptions about the dynamics of the non-cooperative UAV.
Nonetheless, most existing approaches employ only a single sensor modality and assume simplistic kinematic models, most commonly the constant velocity (CV) model \cite{Abdelkader_KF,Pardhasaradhi_KF,Ye_KF}.
Such simplifications may be inadequate for capturing the complex and time-varying behaviors exhibited by non-cooperative UAVs.

With the estimated statistical information of the UAV, robust risk assessment can then be performed.
Signal temporal logic (STL) has emerged as a powerful formal language for specifying and verifying spatiotemporal requirements in cyber-physical systems \cite{Bartocci_STL,Fainekos_STL}.
It is capable of expressing rich requirements, such as reaching goal regions or avoiding restricted zones, and thus it has been extensively applied to verification \cite{Salamati_STL}, control synthesis \cite{Yu_STL,Yu_STL2}, and trace synthesis \cite{Sato_STL}.
For robustness monitoring and risk analysis, the STL robustness was defined to quantify the risk that a stochastic system lacks robustness with respect to an STL specification \cite{Lindemann_STL}.
In practice, temporal logic monitoring can be categorized into offline and online approaches. 
Offline monitoring is applied after complete execution traces have been collected, and its efficiency is proven to depend on the length of execution traces and the size of the formula \cite{Fainekos_STL}.
However, many scenarios require monitoring to be performed during system execution \cite{Deshmukh_STL}, for example, abstract fuel control \cite{Zhang_STL}.
This is especially critical in applications involving non-cooperative UAVs, where offline monitoring could delay threat detection and lead to irreparable losses. 
In such cases, systems often need to set constraints based on real-time perception of the environment or the UAV’s own state, making online monitoring the more suitable approach for these applications.

It should be pointed out that existing STL methods are primarily designed for risk assessment at the present moment and lack the ability to predict future violations, which hinders their effectiveness in early-warning applications.
Concurrently, UAV dynamics are quite complicated and cannot be adequately described by simple CV models, and onboard sensors are often heterogeneous in practice.
To address these gaps in AITSs, this paper proposes a unified monitoring and early warning framework for non-cooperative UAVs which integrates an interacting multiple model multi-rate Kalman filter (IMM-MRKF) with STL.
The key contributions of this work are summarized as follows:
\begin{itemize}
\item 
\textbf{STL-aware adaptive model transition mechanism.} 
We introduce a novel time‑varying Markov transition matrix that incorporates real‑time STL robustness within the proposed IMM‑MRKF.
This mechanism biases the filter toward kinematically plausible and risk‑sensitive motion models, thus effectively embedding formal safety semantics into the state estimation process.
\item 
\textbf{Predictive risk quantification framework via probabilistic reachability.}
Instead of instantaneous robustness evaluation, we derive multi‑step probabilistic reachable sets (PRSs) from the posterior estimate of the IMM‑MRKF.
Meanwhile, we formulate the threshold for future STL robustness, enabling risk assessment over a finite prediction horizon and supporting proactive warning generation for safety management.
\item 
\textbf{Closed-loop unification of estimation and formal verification.}
The proposed framework establishes a principled feedback loop in which STL robustness guides model probabilities, which in turn refine state estimates and future PRS. 
This creates a mathematically coherent pipeline that integrates heterogeneous sensor data, adaptive filtering, and formal verification into a cohesive system for predictive safety assurance.
\end{itemize}

The remainder of this paper is organized as follows. 
Section II presents the system modeling and monitoring problem formulation for non-cooperative UAV.
The design of STL-aware IMM-MRKF is detailed in Section III. 
Then, we introduce STL-based risk analysis approach with robustness semantics and online evaluation in Section IV. 
Moreover, Section V provides experimental results for several components of the proposed algorithm, and Section VI finally concludes the paper with future research directions.
The notations that are frequently used throughout the paper are summarized below.

\emph{Notations}:
$\mathbb{N}$, $\mathbb{R}$, $\mathbb{R}^{n}$, and $\mathbb{R}^{n\times m}$ respectively represent the sets of natural numbers, real numbers, $n$-dimensional real vectors, and $n\times m$ real matrices.
``$\oplus$'' denotes the Minkowski sum for set operations.
The symbols ``$\mathbf{0}_{m}$'', ``$\mathbf{0}_{m \times n}$'', ``$\mathbf{1}_{m}$'', ``$\mathbf{1}_{m \times n}$'' respectively denote the zero matrix, zero vector, all-ones vector, and all-ones matrix with dimension $m$ and $m \times n$, while ``$\mathbf{I}_{n}$'' represents the identity matrix with dimension $n$.
The superscript ``${\mathrm{T}}$'' denotes the matrix transpose.
$\mathrm{diag}\left\{a_{1},\cdots,a_{n}\right\}$ constructs a block diagonal matrix and $\mathrm{col}\left\{a_{1},\cdots,a_{n}\right\}$ forms a column vector whose elements are $a_{1},\cdots,a_{n}$. 
$\mathbb{E}\left\{\cdot\right\}$ means the mathematical expectation, while $\mathbb{P}\left\{\cdot\right\}$ stands for the probability of an event.
$\mathcal{N}(\mu,\Sigma)$ represents a Gaussian distribution with mean $\mu$ and covariance $\Sigma$.

\section{System Model and Problem Formulation}

This section establishes the mathematical foundation for the monitoring and early-warning framework. 
We begin by presenting the dynamical and sensor measurement models for a non-cooperative UAV. Subsequently, we employ STL to provide a formal description of spatial and temporal safety constraints. 
Finally, we synthesize these components to formulate the two core, interconnected problems addressed in this work: accurate state estimation under behavioral uncertainty and predictive, risk-aware safety assessment.

\subsection{UAV Monitoring System} 

We consider the problem of tracking a non-cooperative UAV operating in a two-dimensional surveillance region containing safety-critical restricted areas.
Its kinematic state at discrete time step $k\in\mathbb{N}$ is denoted by $\mathbf{x}_{k}\triangleq[p_{k}^{x};v_{k}^{x};a_{k}^{x};p_{k}^{y};v_{k}^{y};a_{k}^{y}]\in\mathbb{R}^{n_{x}}\ (n_{x}=6)$,
which comprises the planar position, velocity, and acceleration components.
Generally, the state evolution is governed by
\begin{equation}                       
\begin{aligned}
\label{eq:statespace}
    \mathbf{x}_{k+1}
    =
    f(\mathbf{x}_{k},\mathbf{u}_{k})+\mathbf{w}_{k}.
\end{aligned}
\end{equation}
Here, $f$ describes the nominal dynamics, $\mathbf{u}_{k}$ is the unknown control input applied by the non-cooperative agent, and $\mathbf{w}_{k}$ is an additive zero-mean Gaussian process noise vector.

The UAV is assumed to be monitored by a heterogeneous sensor network consisting of radar, vision, and radio-frequency sensing units.
Each sensor $j\in\{1,2,\cdots,N\}$ provides certain measurements at a fixed but different sampling rate $T_{\mathbf{y}_{j}}$.
The measurement from sensor $j$ at its corresponding sampling instant is modeled as:
\begin{equation}                       
\begin{aligned}
\label{eq:measurement}
    \mathbf{y}_{j,T_{\mathbf{y}_{j}}k}
    =
    \mathbf{C}_{j}\mathbf{x}_{T_{\mathbf{y}_{j}}k}+\mathbf{v}_{j,T_{\mathbf{y}_{j}}k}.
\end{aligned}
\end{equation}
In the above, $\mathbf{y}_{j,T_{\mathbf{y}_{j}}k}\in\mathbb{R}^{n_{\mathbf{y}_{j}}}$ represents the measurement at time $T_{\mathbf{y}_{j}}k$ with sampling time $T_{\mathbf{y}_{j}}$.
$\mathbf{C}_{j}\in\mathbb{R}^{n_{\mathbf{y}_{j}}\times n_{x}}$ is the measurement matrix that maps the full state to the observable quantities.
The term $\mathbf{v}_{j,T_{\mathbf{y}_{j}}k}$ represents the sensor measurement noise, also modeled as a zero-mean white Gaussian sequence with covariance $\mathbf{R}_{j}>0$, i.e., $\mathbf{v}_{j,T_{\mathbf{y}_{j}}k}\sim\mathcal{N}(\mathbf{0}_{n_{\mathbf{y}_{j}}},\mathbf{R}_{j})$.
The sampling periods $T_{\mathbf{y}_{j}}$ may be different, hence leading to a multi-rate asynchronous measurement stream.

\subsection{Formal Safety Specifications via STL} 

To precisely encode airspace restrictions, such as static exclusion regions, dynamic safety envelopes, or mission-dependent state constraints, we adopt STL \cite{Zhang_STL}, which is a rigorous formal grammar for specifying spatial and temporal constraints over continuous real-time signals.
Let $\xi=1,\cdots,\Xi$ be the index of a set of $\Xi$ distinct spatial properties or restricted regions.
For the $\xi$-th property, we define an atomic predicate $\mu_{\xi}$ of the form
\begin{equation}                       
\begin{aligned}
\label{eq:mu}
    \mu_{\xi}
    :=
    \theta_{\xi}(\mathbf{x})\geq 0,
\end{aligned}
\end{equation}
where $\theta_{\xi}(\mathbf{x})$ is a scalar, continuously differentiable function whose zero-level set defines the boundary of the region.
Complex tasks or missions can be formally specified using rich STL formulas, which can be constructed by using boolean operators (negation $\neg$, conjunction $\land$) and temporal operators (until $\mathcal{U}$ over a time interval $I=[a,b]$).

Moreover, the robustness degree $\rho(\mathbf{x},\varphi_{\xi},k)$ is a real-valued function that quantifies the margin by which an estimated signal (trajectory) $\hat{\mathbf{x}}$ satisfies the formula $\varphi_{\xi}$.
A positive robustness $\rho(\mathbf{x},\varphi_{\xi},k)>0$ indicates satisfaction, with larger values corresponding to a greater safety margin. 
On the other hand, a negative robustness $\rho(\mathbf{x},\varphi_{\xi},k)<0$ indicates a violation, and its magnitude can be interpreted as a measure of the severity of the violation. 
This quantitative semantics is central to our risk assessment framework.

\subsection{Problem Formulation} 

Integrating the stochastic dynamical model, the multi-sensor measurement model, and the formal safety specifications $\{\varphi_{\xi}\}_{\xi=1}^{\Xi}$, we now formalize the two intertwined problems that constitute the core of this work. 
These problems collectively address the dual challenges of accurate state estimation under motion-mode uncertainty and predictive, risk-aware safety assessment, which are typically treated separately in prior work.

\begin{enumerate}
\item Problem 1 (Adaptive Fusion Estimation):
We first need to design a filtering algorithm at each time step $k$ to produce an accurate estimate $\hat{\mathbf{x}}_{k|k}$ of the UAV's true state $\mathbf{x}_{k}$.
It must fuse the asynchronous and heterogeneous measurement sequences $\{\mathbf{y}_{j,T_{\mathbf{y}_{j}}k}\}_{j=1}^{N}$ from all available sensors.
Meanwhile, it should account for the unknown and time-varying motion mode, such as CV, constant acceleration (CA), and constant turning (CT), of the non-cooperative UAV, and without prior knowledge of its control inputs.
\item Problem 2 (Predictive Risk Assessment):
We need to develop an online method to evaluate the risk of the UAV violating any STL specification $\varphi_{\xi}$ over a finite future horizon.
First, we need to propagate the estimate and uncertainty forward to construct multi-step predictions and their associated probabilistic reachable sets (PRS).
Then, we need to calculate the robustness $\rho(\hat{\mathbf{x}},\varphi_{\xi},\tau)$ based on the known signal $\hat{\mathbf{x}}_{1:k}$ for the formula $\varphi_{\xi}$.
Finally, we implement a multi-layer warning logic that triggers alerts when the predicted risk exceeds predefined thresholds, enabling proactive intervention.
\end{enumerate}

\begin{figure}[htbp]         
\centering
    \includegraphics[width=1.0\columnwidth]{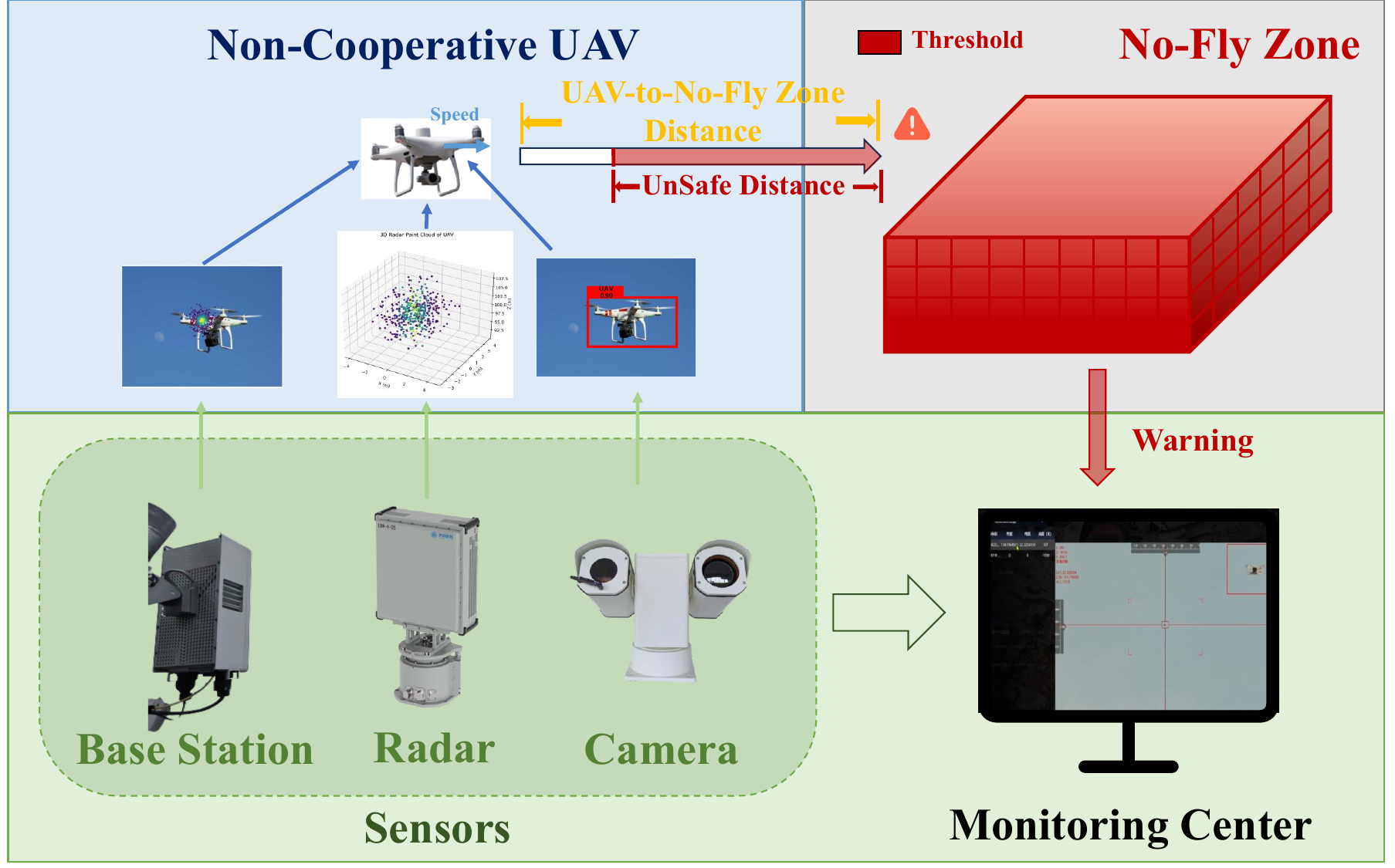}
\caption{
A concept of the proposed monitoring and predictive warning framework for a non-cooperative UAV.
A radar, camera, and base station send measurements to a monitoring center, which estimates the UAV state, evaluates STL-based safety w.r.t. no-fly zones, and issues warnings when unsafe distances are predicted.
}
\label{Fig_Structure}
\end{figure}

The core innovation is embodied in a feedback loop, where the STL robustness signal actively modulates the model transition probabilities of the IMM-MRKF. 
This approach seamlessly integrates high-level safety semantics into the low-level estimation process.
A concept of the proposed monitoring and predictive warning framework for a non-cooperative UAV is shown in Fig. \ref{Fig_Structure}.

\section{STL-Aware Adaptive IMM-MRKF Design} 

Since the control input and exact disturbance statistics are unavailable, the agent dynamics are approximated by a finite set of candidate linear motion models.
To this end, we develop an adaptive IMM-MRKF, where the key innovation is the incorporation of real-time STL robustness evaluations into the filter's core adaptation mechanism. 
This enables the estimator to not only respond to kinematic features but also to proactively adjust its belief about the UAV's motion model based on proximity to restricted zones. 

\subsection{System Representation With State-Space Model}

Since the control input $\mathbf{u}_{k}$ and the statistical information of the noise $\mathbf{w}_{k}$ for a non-cooperative UAV are unknown, we have to approximate the kinematics of the UAV with a finite set of $M$ canonical models.
This set is designed to capture a spectrum of common flight behaviors.
Let $\mathcal{M}_{k}\in\{1,2,\cdots,M\}$ denote the index of the active but unknown motion mode at time $k$.
The switching between these modes is assumed to follow a Markov chain characterized by a transition probability matrix $\Pi_{k|k-1}=[p_{m|i,k|k-1}]_{M\times M}$, where $p_{m|i,k|k-1}=\mathbb{P}(\mathcal{M}_{k}=m|\mathcal{M}_{k-1}=i)$.
For each model $m=1,\dots,M$, the discrete-time state-space representation is described by a linear equation:
\begin{equation}                       
\begin{aligned}
\label{eq:x}
    \mathbf{x}_{k}
    =
    \mathbf{A}_{m}\mathbf{x}_{k-1}+\mathbf{B}_{m}\mathbf{w}_{m,k-1},
\end{aligned}
\end{equation}
where $\mathbf{w}_{m,k}\sim\mathcal{N}(0,\mathbf{Q}_{m})$ is the mode-dependent process noise.

In this work, we employ the following $M$ models to capture the dynamics of the UAV:
\begin{enumerate}
\item ``CV'' model ($m=1$);
\item ``CA'' model ($m=2$);
\item $M-2$ ``CT'' models with different turn rates ($m=3(=CT_{1}),\cdots,m=M(=CT_{M-2})$).
\end{enumerate}
Accordingly, the system parameters $\mathbf{A}_{m}$ and $\mathbf{B}_{m}$ for each model can be described in the following forms:
\begin{equation}                       
\begin{aligned}
\label{eq:A}
    &\mathbf{A}_{1}
    =
    \begin{bmatrix} 
        1 & T_{\mathbf{x}} & 0 & 0 & 0 & 0 \\
        0 & 1 & 0 & 0 & 0 & 0 \\
        0 & 0 & 0 & 0 & 0 & 0 \\
        0 & 0 & 0 & 1 & T_{\mathbf{x}} & 0 \\
        0 & 0 & 0 & 0 & 1 & 0 \\
        0 & 0 & 0 & 0 & 0 & 0
    \end{bmatrix}, 
    \mathbf{B}_{1}
    =
    \begin{bmatrix} 
        \frac{T_{\mathbf{x}}^{2}}{2} & 0 \\
        T_{\mathbf{x}} & 0  \\
        0 & 0  \\
        0 & \frac{T_{\mathbf{x}}^{2}}{2} \\
        0 & T_{\mathbf{x}}  \\
        0 & 0 
    \end{bmatrix}, \\
    &\mathbf{A}_{2}
    =
    \begin{bmatrix} 
        1 & T_{\mathbf{x}} & \frac{T_{\mathbf{x}}^{2}}{2} & 0 & 0 & 0 \\
        0 & 1 & T_{\mathbf{x}}              & 0 & 0 & 0 \\
        0 & 0 & 1              & 0 & 0 & 0 \\
        0 & 0 & 0              & 1 & T_{\mathbf{x}} & \frac{T_{\mathbf{x}}^{2}}{2} \\
        0 & 0 & 0              & 0 & 1 & T_{\mathbf{x}} \\
        0 & 0 & 0              & 0 & 0 & 1
    \end{bmatrix}, 
    \mathbf{B}_{2}
    =
    \begin{bmatrix} 
        \frac{T_{\mathbf{x}}^{2}}{2} & 0 \\
        T_{\mathbf{x}} & 0  \\
        1 & 0  \\
        0 & \frac{T_{\mathbf{x}}^{2}}{2} \\
        0 & T_{\mathbf{x}}  \\
        0 & 1 
    \end{bmatrix}, \\
    &\mathbf{A}_{CT_{i}}
    =
    \begin{bmatrix} 
        1 & \frac{\sin(\omega_{CT_{i}} T_{\mathbf{x}})}{\omega_{CT_{i}}} & 0 & 0 & -\frac{1-\cos(\omega_{CT_{i}} T_{\mathbf{x}})}{\omega_{CT_{i}}} & 0 \\
        0 & \cos(\omega_{CT_{i}} T_{\mathbf{x}})                 & 0 & 0 & -\sin(\omega_{CT_{i}} T_{\mathbf{x}})                  & 0 \\
        0 & 0 & 0 & 0 & 0 & 0 \\
        0 & \frac{1-\cos(\omega_{CT_{i}} T_{\mathbf{x}})}{\omega_{CT_{i}}} & 0 & 1 & \frac{\sin(\omega_{CT_{i}} T_{\mathbf{x}})}{\omega_{CT_{i}}} & 0 \\
        0 & \sin(\omega_{CT_{i}} T_{\mathbf{x}})                 & 0 & 0 & \cos(\omega_{CT_{i}} T_{\mathbf{x}})                  & 0 \\
        0 & 0 & 0 & 0 & 0 & 0
    \end{bmatrix}, \\
    &\mathbf{B}_{CT_{i}}
    =
    \mathbf{B}_{1}\ (\forall\ CT_{i}).
\end{aligned}
\end{equation}

Next, the heterogeneous sensors deliver measurements at asynchronous sampling instants.
To incorporate all available data coherently within the Kalman filter recursion, we employ a centralized fusion approach that constructs an augmented measurement model at each base-rate time step $k$.
Let $\mathbf{y}_{k}^{f}$ denote the augmented vector of all $\Delta$ measurements available at time $k$, $\mathbf{C}_{k}^{f}$ the corresponding augmented observation matrix, and $\mathbf{R}_{k}^{f}$ the associated block-diagonal noise covariance matrix.
After synchronizing under the specified sampling rate of state transition, the valid global measurement can be expressed by:
\begin{equation}                       
\begin{aligned}
\label{eq:y_f}
    &\mathbf{y}_{k}^{f}
    =
    \mathbf{C}_{k}^{f}\mathbf{x}_{k}+\mathbf{v}_{k}^{f},
\end{aligned}
\end{equation}
where
\begin{equation}                       
\begin{aligned}
\label{eq:y_f1}
    \mathbf{y}_{k}^{f}
    =&[\mathbf{y}_{1,k};\cdots;\mathbf{y}_{\Delta,k}]\in\mathbb{R}^{\sum_{j=1}^{\Delta}n_{\mathbf{y}_{j}}}, \\
    \mathbf{C}_{k}^{f}
    =&[\mathbf{C}_{1};\cdots;\mathbf{C}_{\Delta}]\in\mathbb{R}^{\left(\sum_{j=1}^{\Delta}n_{\mathbf{y}_{j}}\right)\times n_{\mathbf{x}}}, \\
    \mathbf{R}_{k}^{f}
    =&
    \mathrm{diag}\{\mathbf{R}_{1},\cdots,\mathbf{R}_{\Delta}\}\in\mathbb{R}^{\left(\sum_{j=1}^{\Delta}n_{\mathbf{y}_{j}}\right)\times\left(\sum_{j=1}^{\Delta}n_{\mathbf{y}_{j}}\right)}.
\end{aligned}
\end{equation}
The block-diagonal structure of $\mathbf{R}_{k}^{f}$ reflects the assumption that inter-sensor correlations are unknown and thus neglected, which is a common simplification in multi-sensor sensing scenarios.


\subsection{STL-Aware and Kinematics-Informed Model Transition}

Notice that the conventional IMM algorithm uses a fixed Markov transition matrix $\Pi$.
Our primary enhancement is to make this matrix time-varying and context-dependent, adapting based on both low-level kinematic cues and high-level safety risks inferred from STL.

Let $\rho(\hat{\mathbf{x}},\varphi_{\xi},\tau)$ denote the robustness of the estimated trajectory $\hat{\mathbf{x}}$ with respect to the STL formula $\varphi_{\xi}$ for the $\xi$-th no-fly zone at time $\tau$.
The risk-aware preference function $\alpha_{m}(\rho(\hat{\mathbf{x}},\varphi_{\xi},\tau))$ is defined as:
\begin{equation}                       
\begin{aligned}
\label{eq:alpha_m}
    \alpha_{m}(\rho(\hat{\mathbf{x}},\varphi_{\xi},\tau))
    = 
    \exp\left(-\kappa_{m}\cdot\max\left(0,1-\frac{\rho(\hat{\mathbf{x}},\varphi_{\xi},\tau)}{\rho_{\mathrm{th}}}\right)\right),
\end{aligned}
\end{equation}
where $\rho_{\mathrm{th}}$ denotes the safe threshold for robustness, and $\kappa_{m}$ is a model-specific sensitivity coefficient.

On the other hand, the instantaneous angular velocity estimate is derived below:
\begin{equation}                       
\begin{aligned}
\label{eq:hat_omega}
    \hat{\omega}_{k}
    =
    \left\{
    \begin{aligned}
        &\hat{\omega}_{\max},\ \mathrm{if}\ \frac{\hat{\omega}_{k}^{o}}{T_{x}}\geq\hat{\omega}_{\max}; \\
        &-\hat{\omega}_{\max},\ \mathrm{if}\ \frac{\hat{\omega}_{k}^{o}}{T_{x}}\leq-\hat{\omega}_{\max}; \\
        &0,\ \mathrm{if}\ \left|\frac{\hat{\omega}_{k}^{o}}{T_{x}}\right|\leq\hat{\omega}_{th}; \\
        &\frac{\hat{\omega}_{k}^{o}}{T_{x}},\ \mathrm{otherwise},
    \end{aligned}
    \right.
\end{aligned}
\end{equation}
where 
\begin{equation}                       
\begin{aligned}
\label{eq:hat_omega_o}
    &\hat{\omega}_{k}^{o}
    =
    ((\gamma_{k}-\gamma_{k-1}+\pi)\mod 2\pi)-\pi, \\
    &\gamma_{k}
    =
    \arctan(\hat{v}_{k|k}^{y}/\hat{v}_{k|k}^{x}).
\end{aligned}
\end{equation}
Then, a kinematic preference function is defined as a Gaussian-shaped weighting:
\begin{equation}                       
\begin{aligned}
\label{eq:beta_m}
    \beta_{m}(\hat{\omega}_{k})
    =
    \exp(-\lambda_{m}(\hat{\omega}_{k}-\overline{\omega}_{m})^{2}),
\end{aligned}
\end{equation}
where $\overline{\omega}_{m}$ is the nominal angular velocity defined as
\begin{equation}                       
\begin{aligned}
\label{eq:bar_omega}
    &\overline{\omega}_{m}
    =
    \left\{
    \begin{aligned}
        &0,\ \mathrm{if}\ m=1; \\
        &0,\ \mathrm{if}\ m=2; \\
        &\omega_{m},\ \mathrm{if}\ m=3,\cdots,M.
    \end{aligned}
    \right.
\end{aligned}
\end{equation}
The parameter $\lambda_{m}$ is the sensitivity coefficient that controls the sensitivity: a large $\lambda_{m}$ imposes a stricter penalty for deviations from $\overline{\omega}_{m}$.

Based on these metrics, the detailed model transition probability $p_{m|i,k|k-1}^{\rho}$ from model $i$ at time $k-1$ to model $m$ at time $k$ is modulated by the kinematic and risk-aware preferences to produce the final time-varying transition probability:
\begin{equation}                       
\begin{aligned}
\label{eq:p_mi_k|k-1^rho}
    p_{m|i,k|k-1}^{\rho}
    =
    \frac{p_{m|i,k|k-1}\alpha_{m}(\rho(\hat{\mathbf{x}},\varphi_{\xi},k))\beta_{m}(\hat{\omega}_{k})}{\sum_{m'=1}^{M}p_{m'|i,k|k-1}\alpha_{m'}(\rho(\hat{\mathbf{x}},\varphi_{\xi},k))\beta_{m'}(\hat{\omega}_{k})},
\end{aligned}
\end{equation}
where $p_{m|i,k|k-1}$ is a nominal (baseline) transition probability.
Thus, the matrix $\Pi_{k|k-1}^{\rho}=[p_{m|i,k|k-1}^{\rho}]_{M\times M}$ effectively implements an STL-aware Markov chain that biases the filter toward behaviorally plausible models in high-risk situations.

\begin{proposition}
Assume that the nominal transition probabilities satisfy $p_{m|i,k|k-1}$ and $\sum_{m=1}^{M}p_{m|i,k|k-1}=1$ for each $i$, and that the preference functions satisfy $\alpha_{m}(\cdot)\geq 0$ and $\beta_{m}(\cdot)\geq 0$ for all $m$.
Then the STL-aware transition probabilities defined in \eqref{eq:p_mi_k|k-1^rho} satisfy
\begin{equation}                       
\begin{aligned}
\label{eq:p_mi_k|k-1^rho>0}
    p_{m|i,k|k-1}\geq 0,\ 
    \sum_{m=1}^{M}p_{m|i,k|k-1}=1,
\end{aligned}
\end{equation}
for each previous mode $i$.
Hence,$\Pi_{k|k-1}^{\rho}$ is a valid time-varying Markov transition matrix.
\end{proposition}
  
Let $\mu_{m,k-1|k-1}=\mathbb{P}(\mathcal{M}_{k-1}=m|\mathbf{Y}_{1:k-1}^{f})$ be the probability of model, given all observations up to time $k-1$ denoted by $\mathbf{Y}_{1:k-1}^{f}=\{\mathbf{y}_{1}^{f},\dots,\mathbf{y}_{k-1}^{f}\}$.
Then, the probability of model $i$ being correct at time $k-1$ given that the model $m$ is correct at time $k$ conditioned on $\mathbf{Y}_{1:k-1}^{f}$ is denoted by $\mu_{i|m,k-1|k}$. 
According to the Bayes rule, we can write it as 
$
\mu_{i|m,k-1|k}
=
\mathbb{P}(\mathcal{M}_{k-1}=i|\mathcal{M}_{k}=m,\mathbf{Y}_{1:k-1}^{f})
=
\frac{\mathbb{P}(\mathcal{M}_{k}=m|\mathcal{M}_{k-1}=i,\mathbf{Y}_{1:k-1}^{f})\mathbb{P}(\mathcal{M}_{k-1}=i|\mathbf{Y}_{1:k-1}^{f})}{\mathbb{P}(\mathcal{M}_{k}=m|\mathbf{Y}_{1:k-1}^{f})}
$.
Therefore, the mixed probability $\mu_{i|m,k-1|k}$ can be calculated by:
\begin{equation}                       
\begin{aligned}
\label{eq:mu_i_m_k-1_k}
    \mu_{i|m,k-1|k}
    =
    \frac{p_{m|i,k|k-1}^{\rho}\mathbf{\mu}_{i,k-1|k-1}}{\bar{c}_{m}},
\end{aligned}
\end{equation}
where $\bar{c}_{m}=\sum_{i=1}^{M}p_{m|i,k|k-1}^{\rho}\mu_{i,k-1|k-1}$ denotes the normalization constant.

The proposed adaptation is introduced at the transition-probability level rather than in the measurement update, so that the Kalman recursion for each candidate model remains unchanged and only the inter-model Bayesian mixing is modified. 
This preserves the computational structure of the standard IMM filter while allowing semantic information to influence mode evolution.

\begin{remark}
If the robustness-aware and kinematics-aware modulation is disabled, e.g., by setting $\kappa_{m}=0$ and $\lambda_{m}=0$ for all $m$, then $\alpha_{m}=1$ and $\beta_{m}=1$.
In this case, \eqref{eq:p_mi_k|k-1^rho} reduces to $p_{m|i,k|k-1}^{\rho}=p_{m|i,k|k-1}$, and the proposed estimator becomes the standard IMM-MRKF.
\end{remark}

\begin{remark}
The STL robustness information is incorporated only through the mode-transition modulation in \eqref{eq:p_mi_k|k-1^rho}. 
The state prediction and measurement update for each model remain identical to those of the standard Kalman filter recursion.
Therefore, the proposed method preserves the recursive computational structure of the IMM estimator, and the additional online computation only arises from robustness evaluation and transition reweighting.
\end{remark}

\subsection{Recursive Algorithm of the STL-Aware IMM-MRKF}

Based on the above knowledge, the proposed IMM-MRKF will operate recursively through the following steps at each discrete time slot $k$.

\textbf{Step 1. Model-Conditioned Re-Initialization:}

Given the previous model probabilities $\mu_{i|m,k-1|k}$ and the estimates $\hat{\mathbf{x}}_{i,k-1|k-1}$, the mixed initial condition for filter $m$ can be computed by:
\begin{equation}                      
\begin{aligned}
\label{eq:hat_x_m0}
    &\hat{\mathbf{x}}_{m,k-1|k-1}^{0}
    =
    \sum_{i=1}^{M}\mu_{i|m,k-1|k}\hat{\mathbf{x}}_{i,k-1|k-1}, \\
    &\mathbf{P}_{m,k-1|k-1}^{0}
    =
    \sum_{i=1}^{M}\mu_{i|m,k-1|k}
    \left(\mathbf{P}_{i,k-1|k-1}\right. \\
    &\left.\qquad\qquad\qquad+\Delta\hat{\mathbf{x}}_{i|m,k-1|k-1}\Delta\hat{\mathbf{x}}_{i|m,k-1|k-1}^{\mathrm{T}}\right),
\end{aligned}
\end{equation}
where $\mu_{i|m,k-1|k}$ is designed in \eqref{eq:mu_i_m_k-1_k} and
\begin{equation}                      
\begin{aligned}
\label{eq:Delta_hat_x_m}
    \Delta\hat{\mathbf{x}}_{i|m,k-1|k-1}
    =
    \hat{\mathbf{x}}_{i,k-1|k-1}-\hat{\mathbf{x}}_{m,k-1|k-1}^{0}.
\end{aligned}
\end{equation}

\textbf{Step 2. Model-Conditioned Prediction:}

Based on the state transition model, the one-step prediction can be computed as follows:
\begin{equation}                      
\begin{aligned}
\label{eq;hat_x_m_k|k-1}
    &\hat{\mathbf{x}}_{m,k|k-1}
    =
    \mathbf{A}_{m}\hat{\mathbf{x}}_{m,k-1|k-1}^{0}, \\
    &\mathbf{P}_{m,k|k-1}
    =
    \mathbf{A}_{m}\mathbf{P}_{m,k-1|k-1}^{0}\mathbf{A}_{m}^{\mathrm{T}}+\mathbf{B}_{m}\mathbf{Q}_{m}\mathbf{B}_{m}^{\mathrm{T}}.
\end{aligned}
\end{equation}

\textbf{Step 3. Augmented Measurement Update:}

For each model $m$, perform a standard Kalman filter step using the augmented measurement model.
At start, the innovation is computed by
\begin{equation}                      
\begin{aligned}
\label{eq:epsilon_m}
    &\boldsymbol{\varepsilon}_{m,k}
    =
    \mathbf{y}_{k}^{f}-\mathbf{C}_{k}^{f}\hat{\mathbf{x}}_{m,k|k-1},
\end{aligned}
\end{equation}
with the covariance
\begin{equation}                      
\begin{aligned}
\label{eq:S_m}
    &\mathbf{S}_{m,k}
    =
    \mathbf{C}_{k}^{f}\mathbf{P}_{m,k|k-1}(\mathbf{C}_{k}^{f})^{\mathrm{T}}+\mathbf{R}_{k}^{f}.
\end{aligned}
\end{equation}

Then, the updated estimate for each model is computed in the following form:
\begin{equation}                      
\begin{aligned}
\label{eq:hat_x_m}
    &\hat{\mathbf{x}}_{m,k|k}
    =
    \hat{\mathbf{x}}_{m,k|k-1}+\mathbf{K}_{m,k}\boldsymbol{\varepsilon}_{m,k}.
\end{aligned}
\end{equation}
The optimal Kalman gain $\mathbf{K}_{m,k}$ is obtained by
\begin{equation}                      
\begin{aligned}
\label{eq:K_m}
    \mathbf{K}_{m,k}
    =
    \mathbf{P}_{m,k|k-1}(\mathbf{C}_{k}^{f})^{\mathrm{T}}\mathbf{S}_{m,k}^{-1},
\end{aligned}
\end{equation}
where 
\begin{equation}                      
\begin{aligned}
\label{eq:P_m}
    &\mathbf{P}_{m,k|k}
    =
    (\mathbf{I}_{n_{x}}-\mathbf{K}_{m,k}\mathbf{C}_{k}^{f})\mathbf{P}_{m,k|k-1}.
\end{aligned}
\end{equation}

\textbf{Step 4. Model Probability Update:}

Based on the Bayes rule, we can derive the relation 
$
\mu_{m,k|k}
=
\mathbb{P}(\mathcal{M}_{k}=m|\mathbf{Y}_{1:k}^{f})
=
\mathbb{P}(\mathcal{M}_{k}=m|\mathbf{y}_{k}^{f},\mathbf{Y}_{1:k-1}^{f})
=
\frac{\mathbb{P}(\mathbf{y}_{k}^{f}|\mathcal{M}_{k}=m)\mathbb{P}(\mathcal{M}_{k}=m|\mathbf{Y}_{1:k-1}^{f})}{\mathbb{P}(\mathbf{y}_{k}^{f}|\mathbf{Y}_{1:k-1}^{f})}
$.
Here, the likelihood function $\Lambda_{m,k}=\mathbb{P}(\mathbf{y}_{k}^{f}|\mathcal{M}_{k}=m)$ for model $m$ can be described by the following form:
\begin{equation}                      
\begin{aligned}
\label{eq:Lambda_m}
    &\Lambda_{m,k}
    =
    \frac{1}{\sqrt{(2\pi)^{\mathbf{dim}(\mathbf{y}_{k}^{f})}|\mathbf{S}_{m,k}|}}\exp(-\frac{1}{2}\boldsymbol{\varepsilon}_{m,k}^{\mathrm{T}}\mathbf{S}_{m,k}^{-1}\boldsymbol{\varepsilon}_{m,k}).
\end{aligned}
\end{equation}
Then, the updated model probabilities $\mu_{m,k|k}$ can be computed by:
\begin{equation}                      
\begin{aligned}
\label{eq:mu_m}
    &\mu_{m,k|k}
    =
    \frac{\Lambda_{m,k}\bar{c}_{m}}{c},
\end{aligned}
\end{equation}
with the normalization constant $c=\sum_{m=1}^{M} \Lambda_{m,k}\bar{c}_{m}$.

\textbf{Step 5. Combined Estimate Fusion:}

The final state estimate and covariance are computed as a weighted sum over all models:
\begin{equation}                      
\begin{aligned}
\label{eq:hat_x}
    &\hat{\mathbf{x}}_{k|k}
    =
    \sum_{m=1}^{M}\mu_{m,k|k}\hat{\mathbf{x}}_{m,k|k}, \\
    &\mathbf{P}_{k|k}
    =
    \sum_{m=1}^{M}\mu_{m,k|k}
    \left(\mathbf{P}_{m,k|k}+\Delta\hat{\mathbf{x}}_{m,k|k}\Delta\hat{\mathbf{x}}_{m,k|k}^{\mathrm{T}}\right),
\end{aligned}
\end{equation}
with
\begin{equation}                      
\begin{aligned}
\label{hat_x}
    \Delta\hat{\mathbf{x}}_{m,k|k}
    =
    \hat{\mathbf{x}}_{m,k|k}-\hat{\mathbf{x}}_{k|k}.
\end{aligned}
\end{equation}

The complete online monitoring procedure at each time step $k$ is summarized in Algorithm \ref{alg:STL_IMM_MRKF}.

\begin{algorithm}[htbp]
\caption{Monitoring with STL-Aware IMM-MRKF}
\label{alg:STL_IMM_MRKF}
\begin{algorithmic}[1]
    \REQUIRE 
    \STATE Previous combined state estimate $\hat{\mathbf{x}}_{k-1|k-1}$ and covariance $\mathbf{P}_{k-1|k-1}$ and model-conditioned estimates $\{\hat{\mathbf{x}}_{i,k-1|k-1}, \mathbf{P}_{i,k-1|k-1}\}_{i=1}^{M}$;
    \STATE Previous model probabilities $\{\mu_{m,k-1|k-1}\}_{m=1}^{M}$;
    \STATE Asynchronous, heterogeneous measurements $\mathbf{y}_{k}^{f}$ and corresponding matrices $\mathbf{C}_{k}^{f}, \mathbf{R}_{k}^{f}$;
    \STATE STL robustness $\rho(\hat{\mathbf{x}},\varphi_\xi,k-1)$ for no-fly zones;
    \STATE Model set $\{ \mathbf{A}_m, \mathbf{B}_m, \mathbf{Q}_m \}_{m=1}^M$;
    \STATE STL-aware transition parameters $\{ \kappa_m, \lambda_m \}_{m=1}^M$ and threshold $\rho_{\mathrm{th}}$.
    
    \ENSURE Updated state estimate $\hat{\mathbf{x}}_{k|k}$, covariance $\mathbf{P}_{k|k}$, and model probabilities $\{\mu_{m,k|k}\}_{m=1}^{M}$.
    
    \FOR{$m = 1$ to $M$}
    
        \STATE \textbf{Step 1: Model-Conditioned Re-Initialization}
        \STATE Compute risk-aware preference $\alpha_{m}(\rho(\hat{\mathbf{x}},\varphi_{\xi},k))$ and kinematic preference $\beta_m(\hat{\omega}_k)$ using \eqref{eq:alpha_m} and \eqref{eq:beta_m};
        \STATE Compute $p^{\rho}_{m|i,k|k-1}$ via \eqref{eq:p_mi_k|k-1^rho};
        \STATE Compute probability $\mu_{i|m,k-1|k}$ using \eqref{eq:mu_i_m_k-1_k};
        \STATE Compute $\hat{\mathbf{x}}_{m,k-1|k-1}^{0}$ using \eqref{eq:hat_x_m0};
    
        \STATE \textbf{Step 2: Model-Conditioned Prediction}
        \STATE Compute $\hat{\mathbf{x}}_{m,k|k-1}$ using \eqref{eq;hat_x_m_k|k-1};
    
        \STATE \textbf{Step 3: Augmented Measurement Update}
        \STATE Compute $\hat{\mathbf{x}}_{m,k|k}$ using \eqref{eq:hat_x_m};
    
        \STATE \textbf{Step 4: Model Probability Update}
        \STATE Compute $\mu_{m,k|k}$ using \eqref{eq:mu_m};
        
    \ENDFOR
    
    \STATE \textbf{Step 5: Combined Estimate Fusion}
    \STATE Compute $\hat{\mathbf{x}}_{k|k}$ using \eqref{eq:hat_x};
    
    \STATE \textbf{Output:} $\hat{\mathbf{x}}_{k|k}$, $\mathbf{P}_{k|k}$, $\{ \mu_{m,k|k} \}_{m=1}^M$.
\end{algorithmic}
\end{algorithm}


\begin{remark}
Compared with the standard IMM-MRKF, the additional computation of the proposed method consists of:
1) evaluating the robustness-dependent and kinematics-dependent weights  and $\alpha_{m}=1$ and $\beta_{m}=1$ for $m=1,\cdots,M$.
2) normalizing the resulting transition probabilities in (13); and
3) propagating finite-horizon predicted moments for robustness evaluation.
Hence, the extra online complexity is mainly $O(M^{2})$ for transition modulation and $O(M\delta)$ for horizon-$\delta$prediction, in addition to the standard IMM filtering cost.
\end{remark}

\section{STL-Based Risk Analysis for Predictive Monitoring} 

Building upon the adaptive state estimation framework established in Section III, this section presents the methodology for predictive risk assessment and early warning generation.
The main objective is to leverage the stochastic state estimates produced by the IMM-MRKF, which contain the most likely trajectory and the associated uncertainty, to formally evaluate future compliance with the spatio-temporal safety constraints encoded in STL.
In the following parts, the parameters with subscription ``$\tau|\tau$'' correspond to $\tau\in[0,k]$ and ``$\tau|k$'' for $\tau\in(k,k+\delta]$ with $\delta>0$.

\subsection{Multi-Step Prediction and Probabilistic Reachable Sets}

To anticipate potential violations before they occur, we first propagate the current state estimate and its uncertainty distribution to the time $\tau\in(k,k+\delta]$ over a finite prediction horizon of $\delta>0$ discrete time steps.
This propagation must account for the multi-modal nature of the state estimate provided by the IMM-MRKF and the time-varying model transition process.

The evolution of the model probabilities over the horizon is governed by the STL-aware Markov transition matrix $\Pi_{k|k-1}^{\rho}$.
According to the law of total probability, the model probability can be described by
$
\mu_{m,\tau|k}
=
\mathbb{P}(\mathcal{M}_{\tau}=m|\mathbf{Y}_{1:k}^{f})
\ (\forall\ \tau\in(k,k+\delta])
$.
Since the model-switching process is Markovian, $\mathcal{M}_{k}$ is irrelevant to measurements, and one has 
$
\mathbb{P}(\mathcal{M}_{\tau}=m|\mathbf{Y}_{1:k}^{f})
=
\sum_{i=1}^{M}\mathbb{P}(\mathcal{M}_{\tau}=m|\mathcal{M}_{k}=i)\mathbb{P}(\mathcal{M}_{k}=i|\mathbf{Y}_{1:k}^{f})
$.
In this case, the predictive model probability $\mu_{m,\tau|k}$ is derived as
\begin{equation}                      
\begin{aligned}
\label{eq:mu_m,tau|k}
    \mu_{m,\tau|k}
    =
    \sum_{i=1}^{M}[\Pi_{\tau|k}^{\rho}]_{m,i}\mu_{i,k|k},
\end{aligned}
\end{equation}
where $[\Pi_{\tau|k}^{\rho}]_{m,i}$ means the $(m,i)$-entry of the matrix $\Pi_{\tau|k}^{\rho}$ that is
\begin{equation}                      
\begin{aligned}
\label{eq:Pi_tau|k}
    \Pi_{\tau|k}^{\rho}
    =
    \Pi_{\tau|\tau-1}^{\rho}\cdots\Pi_{k+1|k}^{\rho}.
\end{aligned}
\end{equation}
In practice, for computational tractability over finite horizons, the time-varying matrix $\Pi_{k+h|k+h-1}^{\rho}\ (h\in[1,\tau-k])$ may be assumed constant over the prediction window, i.e., $\Pi_{k+h|k+h-1}^{\rho}\approx\Pi_{k|k-1}^{\rho}\ (\forall\ h\in[1,\tau-k])$.

For each candidate model $m$, the $(\tau-k)$-step state prediction and its associated covariance, starting from the mixed initial condition $(\hat{\mathbf{x}}_{m,k|k}^{0},\mathbf{P}_{m,k|k}^{0})$, are given by:
\begin{equation}                      
\begin{aligned}
\label{eq:hat_x_m,tau|k}
    &\hat{\mathbf{x}}_{m,\tau|k}
    =
    \mathbf{A}_{m}^{\tau-k}\hat{\mathbf{x}}_{m,k|k}^{0}, \\
    &\mathbf{P}_{m,\tau|k}
    =
    \mathbf{A}_{m}^{\tau-k}\mathbf{P}_{m,k|k}^{0}(\mathbf{A}_{m}^{\tau-k})^{\mathrm{T}} \\
    &\qquad\qquad+
    \sum_{t=0}^{\tau-k-1}\mathbf{A}_{m}^{t}\mathbf{B}_{m}\mathbf{Q}_{m}\mathbf{B}_{m}^{\mathrm{T}}(\mathbf{A}_{m}^{\mathrm{T}})^{t}.
\end{aligned}
\end{equation}
Then, the combined predicted state estimate and covariance are obtained by merging across all models:
\begin{equation}                      
\begin{aligned}
\label{eq:hat_x_k+tau|K}
    &\hat{\mathbf{x}}_{\tau|k}
    =
    \sum_{m=1}^{M}\mu_{m,\tau|k}\hat{\mathbf{x}}_{m,\tau|k}, \\
    &\mathbf{P}_{\tau|k}
    =
    \sum_{m=1}^{M}\mu_{m,\tau|k}\left(\mathbf{P}_{m,\tau|k}+\Delta\hat{\mathbf{x}}_{m,\tau|k}\Delta\hat{\mathbf{x}}_{m,\tau|k}^{\mathrm{T}}\right),
\end{aligned}
\end{equation}
with
\begin{equation}                      
\begin{aligned}
\label{eq:Delta_hat_x}
    \Delta\hat{\mathbf{x}}_{m,\tau|k}
    =
    \hat{\mathbf{x}}_{m,\tau|k}-\hat{\mathbf{x}}_{\tau|k}.
\end{aligned}
\end{equation}
Note that the IMM posterior is in general a Gaussian mixture. 
Here, we adopt the standard moment-matching approximation and represent it by a single Gaussian distribution, whose mean and covariance are computed as above.
This combined distribution $\mathbf{x}_{\tau|k}\sim(\hat{\mathbf{x}}_{\tau|k},\mathbf{P}_{\tau|k})$ approximates the future state uncertainty.

Next, to rigorously reason about the uncertainty in these predictions, we introduce the concept of a PRS, which provides a confidence region for the UAV's future state.
The detailed definition of $\zeta$-probabilistic reachable set ($\zeta$-PRS) is given as follows.

\begin{definition}[\textbf{$\zeta$-Probabilistic Reachable Set}]
Let $\mathbf{x}_{\tau|k}$ denote a random state variable at time $k$, conditioned on information available up to time $s$, where $s\leq\tau$. 
A set $\mathcal{X}_{\tau|s}^{\zeta}\subseteq\mathbb{R}^{n_{x}}$ is called a $\zeta$-PRS of $\mathbf{x}_{\tau|s}$ at probability level $1-\zeta$ (where $\zeta\in(0,1)$) if
\begin{equation}                       
\begin{aligned}
\label{eq:def_PRS}
    \mathbb{P}(\mathbf{x}_{\tau|s}\in\mathcal{X}_{\tau|s}^{\zeta})
    \geq
    1-\zeta.
\end{aligned}
\end{equation}
\end{definition}

Given that our combined state estimate is approximately Gaussian, i.e., $\mathbf{x}_{\tau|\tau}\sim\mathcal{N}(\hat{\mathbf{x}}_{\tau|\tau},\mathbf{P}_{\tau|\tau})\ (\forall\ \tau\in[0,k])$, an ellipsoidal representation of the $\zeta$-PRS is both compact and interpretable, which can be defined as:
\begin{equation}                       
\begin{aligned}
\label{eq:PRS1}
    &\mathcal{X}_{\tau|\tau}^{\zeta}
    \triangleq
    \{\mathbf{x}\in\mathbb{R}^{n_{x}},\forall\ \tau\in[0,k]
    | \\
    &\qquad\qquad(\mathbf{x}-\hat{\mathbf{x}}_{\tau|\tau})^{\mathrm{T}}\mathbf{P}_{\tau|\tau}^{-1}(\mathbf{x}-\hat{\mathbf{x}}_{\tau|\tau})\leq\chi_{n_{x}}^{2}(1-\zeta)\},
\end{aligned}
\end{equation}
where $\chi_{n_{x}}^{2}(1-\zeta)$ is the quantile of the chi-squared distribution with $n_{x}$ degrees of freedom.
Analogously, the future $\zeta$-PRS can be obtained for the prediction horizon $\delta>0$ as:
\begin{equation}                       
\begin{aligned}
\label{eq:PRS2}
    &\mathcal{X}_{\tau|k}^{\zeta}
    \triangleq
    \{\mathbf{x}\in\mathbb{R}^{n_{x}},\forall\ \tau\in(k,k+\delta]
    | \\
    &\qquad\qquad(\mathbf{x}-\hat{\mathbf{x}}_{\tau|k})^{\mathrm{T}}\mathbf{P}_{\tau|k}^{-1}(\mathbf{x}-\hat{\mathbf{x}}_{\tau|k})
    \leq\chi_{n_{x}}^{2}(1-\zeta)\}.
\end{aligned}
\end{equation}
This ellipsoid defines a region in the state space where the UAV is expected to reside at future time $\tau$ with probability at least $1-\zeta$.
The sequence $\{\mathcal{X}_{\tau|k}^{\zeta}\}_{\tau\in(k,k+\delta]}$ forms a probabilistic tube that encapsulates the future trajectory uncertainty.

\subsection{STL Specifications and Robustness Semantics}

The STL robustness semantics defined in Section II-B are deterministic, evaluating a specific signal trace. 
For risk assessment, we must evaluate the robustness of a distribution, namely, the likelihood and severity with which the random future state $\mathbf{x}_{k+\delta|k}$ might violate a specification.
To bridge this gap, we extend the robustness semantics to account for the uncertainty represented by the PRS.
The concrete STL syntax and semantics are proposed below by considering our risk assessment context.


\begin{definition}[\textbf{STL Syntax}]
Let $\xi$ index a set of no-fly zones or spatial properties.
The STL formula $\varphi_{\xi}$ for the $\xi$-th property is defined using the Backus-Naur form:
\begin{equation}                       
\begin{aligned}
\label{eq:STL_Syntax}
    \varphi_{\xi}
    :=
    &\mu_{\xi} \ |\ 
    \neg\varphi_{\xi} \ |\ 
    \varphi_{\xi,1}\land\varphi_{\xi,2} \ |\ 
    \varphi_{\xi,1}\mathcal{U}_{I}\varphi_{\xi,2},
\end{aligned}
\end{equation}
where $\mu_{\xi}:=\theta_{\xi}(\mathbf{x})\geq 0$ is the atomic predicate with the predicate function $\theta_{\xi}(x)$.
$\neg$, $\land$, and $\mathcal{U}$ are basic operators that respectively mean negation, conjunction, and until.
Other operators can be derived by using these operators, such as $\lor=\neg\land$, $\mathcal{G}_{I}\varphi=\neg\mathcal{F}_{I}\neg\varphi$, and $\mathcal{F}_{I}\varphi=\top\mathcal{U}_{I}\varphi$.
$I=[a,b]\subseteq\mathbb{R}_{\geq0}$ is the time interval.
\end{definition}

An observation map $\mathcal{O}^{\mu_{\xi}}$ with a predicate $\mu_{\xi}$ can be constructed to indicate regions where the predicate $\mu_{\xi}$ is true, i.e.,
\begin{equation}                       
\begin{aligned}
\label{eq:O_mu_xi}
    \mathcal{O}^{\mu_{\xi}}
    :=
    \mu_{\xi}^{-1}(\top),
\end{aligned}
\end{equation}
where $\mu_{\xi}^{-1}(\top)$ denotes the inverse image of $\top$ under the function $\mu_{\xi}$.
Also, we define the satisfaction function $\beta(\hat{\mathbf{x}},\varphi_{\xi},k)$, where $\beta(\hat{\mathbf{x}},\varphi_{\xi},k)=\top$ indicates that the signal $\hat{\mathbf{x}}$ satisfies the formula $\varphi_{\xi}$ at time $k$.

Importantly, it is also of interest to quantify the degree of satisfaction, that is to assess how robustly the estimate $\hat{\mathbf{x}}$ satisfies the STL formula $\varphi_{\xi}$.
For the $\xi$-th no-fly zone, let us first define the set of signals that violate its formula $\varphi_{\xi}$:
\begin{equation}                       
\begin{aligned}
\label{eq:L_neg_phi_xi}
    \mathcal{L}^{\neg\varphi_{\xi}}_{\tau}
    :=
    \{\hat{\mathbf{x}}|\beta(\hat{\mathbf{x}},\neg\varphi_{\xi},\tau)=\top\}.
\end{aligned}
\end{equation}
To measure distances between the signal $\hat{\mathbf{x}}$ and a two-dimensional trajectory $\mathbf{p}$, let us define the metric
\begin{equation}                       
\begin{aligned}
\label{eq:kappa}
    \kappa(\hat{\mathbf{x}},\mathbf{p})
    =
    \sup_{t\in[0,k]}\ \mathrm{dist}(\mathbf{C}_{pos}\hat{\mathbf{x}}_{t|t},\mathbf{p}_{t}), \\
\end{aligned}
\end{equation}
where 
\begin{equation}                       
\begin{aligned}
\label{eq:dist}
    &\mathrm{dist}(\mathbf{C}_{pos}\hat{\mathbf{x}}_{t|t},\mathbf{p}_{t})
    =
    \|\mathbf{C}_{pos}\hat{\mathbf{x}}_{t|t}-\mathbf{p}_{t}\|_{l}, \\
    &\mathbf{C}_{pos}
    \triangleq
    \begin{bmatrix}
    1 & 0 & 0 & 0 & 0 & 0 \\
    0 & 0 & 0 & 1 & 0 & 0
    \end{bmatrix}.
\end{aligned}
\end{equation}
More generally, for a metric space $\mathbf{p}\in\mathcal{P}$, the distance between a trajectory or a position and a space can be written as
\begin{equation}                       
\begin{aligned}
\label{eq:bar_kappa}
    &\overline{\kappa}(\hat{\mathbf{x}},\mathcal{P})
    =
    \inf_{\mathbf{p}\in\mathcal{P}} \kappa(\hat{\mathbf{x}},\mathbf{p}), \\
    &\overline{\mathrm{dist}}(\hat{\mathbf{x}}_{t|t},\mathcal{P}_{t})
    =
    \inf_{\mathbf{p}_{t}\in\mathcal{P}_{t}} \|\mathbf{C}_{pos}\hat{\mathbf{x}}_{t|t}-\mathbf{p}_{t}\|_{2}.
\end{aligned}
\end{equation}

By combining the violation space $\mathrm{cl}(\mathcal{L}^{\neg\varphi_{\xi}}_{\tau})$, the robustness degree (RD) can be described as the following form:
\begin{equation}                       
\begin{aligned}
\label{eq:1}
    RD^{\varphi_{\xi}}_{\tau}(\hat{\mathbf{x}})
    =
    \overline{\kappa}(\hat{\mathbf{x}},\mathrm{cl}(\mathcal{L}^{\neg\varphi_{\xi}}_{\tau})),
\end{aligned}
\end{equation}
where $\mathrm{cl}(\mathcal{L}^{\neg\varphi_{\xi}}_{\tau})$ denotes the closure of the set $\mathcal{L}^{\neg\varphi_{\xi}}_{\tau}$.
However, calculating the RD can be challenging in practice since the set $\mathcal{L}^{\neg\varphi_{\xi}}_{\tau}$ is hard to calculate.
Therefore, the robustness semantics are required for approximation.
Given the constraints present in real-world scenarios, we assume that the robustness semantic is bounded i.e.,$\rho(\hat{\mathbf{x}},\mu_{\xi},\tau)\in[\rho_{\min},\rho_{\max}]\ (\forall\ \tau\in[0,k+\delta])$.
In practice, $\rho_{\max}$ and $\rho_{\min}$ are chosen according to physically plausible distance bounds in the given monitoring area, for example, $\rho_{\max}=d_{\max},\rho_{\min}=-d_{\max}$.
Then, the recursions for the robustness semantics are given as follows.

\begin{definition}[\textbf{STL Robustness Semantics}]
For the state trajectory $\hat{\mathbf{x}}$, time instant $\tau$, and the STL formula $\varphi_{\xi}$ for $\xi$-th zone, the robustness $\rho(\hat{\mathbf{x}},\varphi_{\xi},\tau)$ is defined by induction on the construction of formulas as follows:
\begin{equation}                       
\begin{aligned}
\label{eq:STL_Robustness_Semantic}
    &\rho(\hat{\mathbf{x}},\mu_{\xi},\tau) \\
    &=
    \left\{
    \begin{aligned}
        &\overline{\mathrm{dist}}(\hat{\mathbf{x}}_{\tau|\tau},\mathrm{cl}(\overline{\mathcal{O}}_{\tau|\tau}^{\neg\mu_{\xi}})),\ \mathrm{if}\ \hat{\mathbf{x}}_{\tau|\tau}\in\overline{\mathcal{O}}_{\tau|\tau}^{\mu_{\xi}}\ \&\ \tau\in[0,k]; \\
        &\overline{\mathrm{dist}}(\hat{\mathbf{x}}_{\tau|k},\mathrm{cl}(\overline{\mathcal{O}}_{\tau|k}^{\neg\mu_{\xi}})),\ \mathrm{if}\ \hat{\mathbf{x}}_{\tau|k}\in\overline{\mathcal{O}}_{\tau|k}^{\mu_{\xi}}\ \&\ \tau\in(k,k+\delta]; \\
        &-\overline{\mathrm{dist}}(\hat{\mathbf{x}}_{\tau|\tau},\mathrm{cl}(\overline{\mathcal{O}}_{\tau|\tau}^{\mu_{\xi}})),\ \mathrm{if}\ \hat{\mathbf{x}}_{\tau|\tau}\in\overline{\mathcal{O}}_{\tau|\tau}^{\neg\mu_{\xi}}\ \&\ \tau\in[0,k]; \\
        &-\overline{\mathrm{dist}}(\hat{\mathbf{x}}_{\tau|k},\mathrm{cl}(\overline{\mathcal{O}}_{\tau|k}^{\mu_{\xi}})),\ \mathrm{if}\ \hat{\mathbf{x}}_{\tau|k}\in\overline{\mathcal{O}}_{\tau|k}^{\neg\mu_{\xi}}\ \&\ \tau\in(k,k+\delta]; \\
        &\rho_{\max},\ \mathrm{otherwise},
    \end{aligned}
    \right. \\
    &\rho(\hat{\mathbf{x}},\neg\varphi_{\xi},\tau)
    =
    -\rho(\hat{\mathbf{x}},\varphi_{\xi},\tau), \\
    &\rho(\hat{\mathbf{x}},\varphi_{\xi,1}\land\varphi_{\xi,2},\tau)
    =
    \min\{\rho(\hat{\mathbf{x}},\varphi_{\xi,1},\tau),\rho(\hat{\mathbf{x}},\varphi_{\xi,2},\tau)\}, \\
    &\rho(\hat{\mathbf{x}},\varphi_{\xi,1}\mathcal{U}_{I}\varphi_{\xi,2},\tau)
    =
    \sup_{\tau'\in\tau\oplus I}\left\{\min\{\rho(\hat{\mathbf{x}},\varphi_{\xi,2},\tau')\},\right. \\
    &\qquad\qquad\qquad\qquad\qquad\left.\inf_{\tau''\in[\tau,\tau')}\{\rho(\hat{\mathbf{x}},\varphi_{\xi,1},\tau'')\}\right\},
\end{aligned}
\end{equation}
where 
\begin{equation}                       
\begin{aligned}
\label{eq:O}
    \overline{\mathcal{O}}_{\tau|\tau}^{\neg\mu_{\xi}}
    \triangleq&
    \mathcal{O}^{\neg\mu_{\xi}}\oplus\mathcal{X}_{\tau|\tau}^{\zeta}\ (\tau\in[0,k]), \\
    \overline{\mathcal{O}}_{\tau|k}^{\neg\mu_{\xi}}
    \triangleq&
    \mathcal{O}^{\neg\mu_{\xi}}\oplus\mathcal{X}_{\tau|k}^{\zeta}\ (\tau\in(k,k+\delta]).
\end{aligned}
\end{equation}
\end{definition}

\subsection{Online Risk Analysis}

Building upon the robustness semantics defined above, we will translate the robustness into a time-varying risk score and a corresponding staged warning system.
For each no-fly zone, we define its risk metric as a function of the robustness:
\begin{equation}                       
\begin{aligned}
\label{eq:R}
    R_{\tau}
    =
    \max\left(0,1-\frac{\rho(\hat{\mathbf{x}},\varphi_{\xi},\tau)}{\rho_{\mathrm{th}}}\right)\ (\forall\ \xi),
\end{aligned}
\end{equation}
where $\rho_{\mathrm{th}}(<\rho_{\max})$ stands for the threshold.
The risk level is normalized to the range $[0,1]$:
\begin{itemize}
\item $R_{\tau}=0$ when $\rho(\hat{\mathbf{x}},\varphi_{\xi},\tau)\geq\rho_{\mathrm{th}}$, indicating no anticipated risk within the confidence region.
\item $R_{\tau}\in(0,1)$ means a potential violation where the severity is quantified by the ratio of the robustness margin to the threshold.
\item $R_{\tau}=1$ represents a severe and certain future violation.
\end{itemize}

To facilitate operational decision-making, the continuous risk metric is mapped to a discrete set of warning levels $W_{\tau}$.
The system issues different $W_{\tau}$ according to the following logic:
\begin{itemize}
\item $W_{\tau}(=0)\gets\mathrm{Safe}$:
Triggered if $R_{\tau}\in[0,0.5)$.
\item $W_{\tau}(=1)\gets\mathrm{Advisory\ Warning}$:
Triggered if $R_{\tau}\in[0.5,0.7)$.
A potential violation is plausible in the prediction horizon. 
The monitoring system will recommend increased surveillance and prepares contingency plans.
\item $W_{\tau}(=2)\gets\mathrm{Alert\ Warning}$:
Triggered if $R_{\tau}\in[0.7,0.9)$.
A violation is likely. 
The monitoring system will generate a formal alert to automatic controllers or human operators and suggest preparatory intervention actions.
\item $W_{\tau}(=3)\gets\mathrm{Critical\ Warning}$:
Triggered if $R_{\tau}\in[0.9,1)$.
A violation is highly imminent and severe. 
The monitoring system will trigger the highest priority response protocols, which may include automated initiation of countermeasures or immediate emergency alerts.
\end{itemize}

The complete online risk analysis procedure at each time step $k$ is summarized in Algorithm \ref{alg:STL_risk_warning}.
It integrates the STL-aware IMM-MRKF estimation with the predictive reachability and risk evaluation into a cohesive monitoring loop.
It should be noted that the numerical breakpoints or threshold parameters are user-selected and can be tuned for different operational policies, without affecting the underlying monitoring and risk-evaluation structure.

\begin{algorithm}[htbp]
\caption{Risk Assessment and Early Warning}
\label{alg:STL_risk_warning}
\begin{algorithmic}[1]
    \REQUIRE 
    \STATE Current state estimate $\hat{\mathbf{x}}_{k|k}$ and covariance $\mathbf{P}_{k|k}$;
    \STATE Model probabilities $\{\mu_{m,k|k}\}_{m=1}^{M}$;
    \STATE Model transition probabilities $\Pi_{k|k-1}^{\rho}$;
    \STATE STL formulas $\{\varphi_{\xi}\}_{\xi=1}^{\Xi}$ for no-fly zones;
    \STATE Safety threshold $\rho_{\text{th}}$, prediction horizon $\delta$;
    \STATE Confidence level $\zeta$ for PRS.
    
    \ENSURE Risk metric $R_{\tau}$ and warning level $W_{\tau}$.
    
    \STATE \textbf{Initialize:} $R_{\tau}=0$, $W_{\tau}(=0)\gets\mathrm{Safe}$.
    
    \FOR{each no-fly zone $\xi = 1$ to $\Xi$}
    
        \STATE Compute $\hat{\mathbf{x}}_{\tau|k},\mathbf{P}_{\tau|k}\ (\forall\ \tau\in(k,k+\delta])$ using \eqref{eq:hat_x_m,tau|k};
        \STATE Construct $\zeta$-PRS $\mathcal{X}^\zeta_{\tau|\tau}\ (\forall\ \tau\in[0,k])$ and $\mathcal{X}^\zeta_{\tau|k}\ (\forall\ \tau\in(k,k+\delta])$ using \eqref{eq:PRS1} and \eqref{eq:PRS2}.
        \STATE Compute $\overline{\mathcal{O}}_{\tau|\tau}^{\neg\mu_{\xi}}$ and $\overline{\mathcal{O}}_{\tau|k}^{\neg\mu_{\xi}}$ using \eqref{eq:O}.
        
        
        \STATE Compute the robustness $\rho(\hat{\mathbf{x}},\varphi_{\xi},\tau)$ using \eqref{eq:STL_Robustness_Semantic}.
        
    \ENDFOR
        
    \STATE Compute $R_{\tau}=\max\left(0,1-\frac{\rho(\hat{\mathbf{x}},\varphi_{\xi},\tau)}{\rho_{\mathrm{th}}}\right)\ (\forall\ \xi)$ using \eqref{eq:R};
    
    \IF{$R_{\tau}\in[0,0.5)$}
        \STATE $W_{\tau}(=0)\gets\mathrm{Safe}$;
    \ELSIF{$R_{\tau}\in[0.5,0.7)$}
        \STATE $W_{\tau}(=1)\gets\mathrm{Advisory\ Warning}$;
    \ELSIF{$R_{\tau}\in[0.7,0.9)$}
        \STATE $W_{\tau}(=2)\gets\mathrm{Alert\ Warning}$;
    \ELSE
        \STATE $W_{\tau}(=3)\gets\mathrm{Critical\ Warning}$.
    \ENDIF
    
    
    \STATE \textbf{Output:} Risk metric $R_{\tau}$, warning level $W_{\tau}$.
\end{algorithmic}
\end{algorithm}

\section{Experimental Results}

This section presents a comprehensive experimental evaluation of the proposed STL-aware IMM-MRKF framework for monitoring a non-cooperative UAV.
The primary objectives are to validate: 
(i) the accuracy and robustness of state estimation under multi-rate sensing; 
(ii) the efficacy of the STL-aware model adaptation mechanism; 
(iii) the predictive capability of the risk assessment for timely warnings. 

\subsection{Experimental Setup}

\begin{figure}[htbp]         
\centering
    \includegraphics[width=1.0\columnwidth]{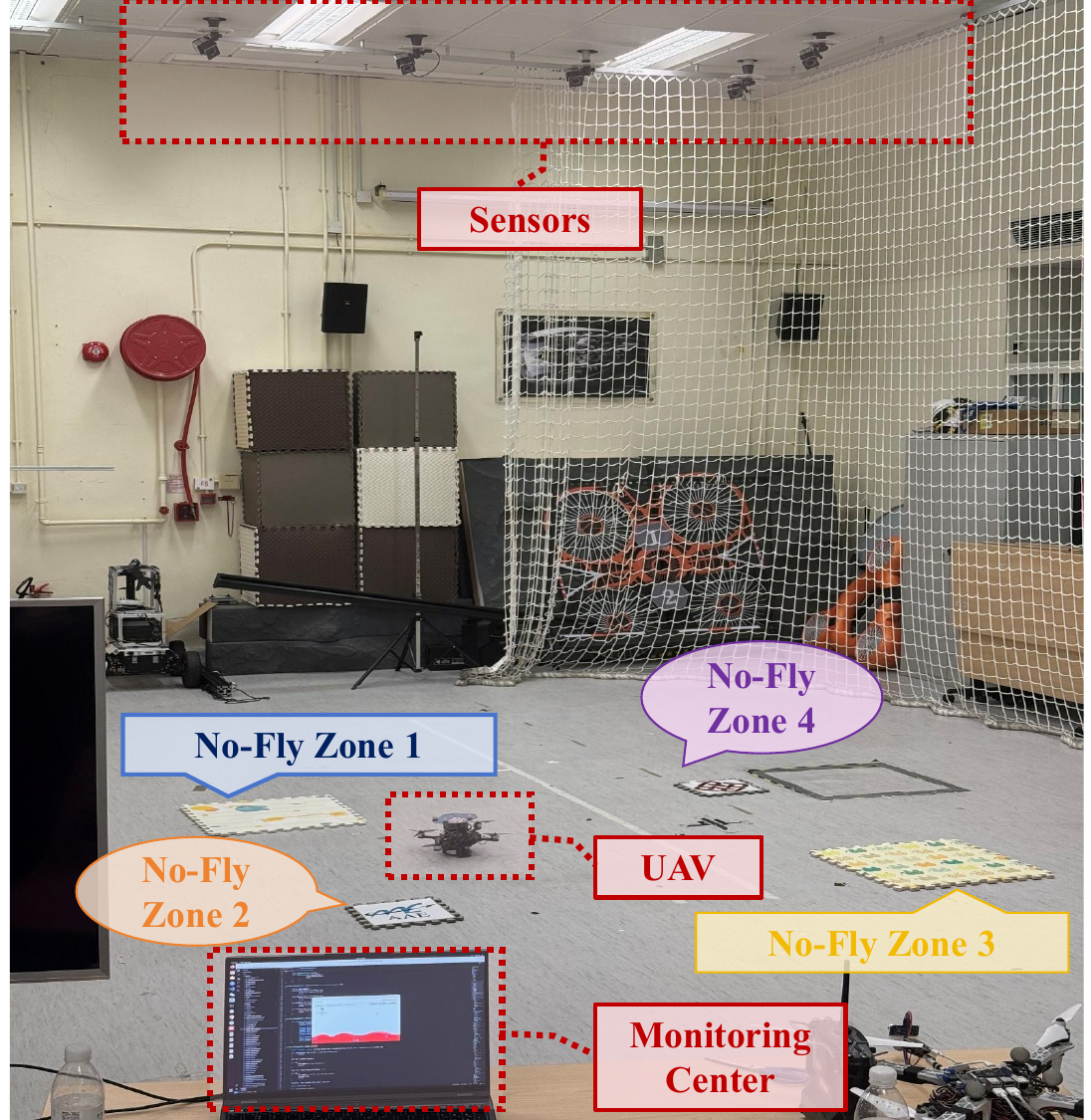}
    \caption{
    Experimental setup for monitoring a non-cooperative UAV in an indoor environment emulating an urban low-altitude airspace. 
    The UAV follows a predefined delivery path while four static no-fly zones are defined around sensitive areas. 
    A Vicon motion capture system provides ground truth, from which three heterogeneous virtual sensors are simulated and used as inputs to the proposed STL-aware IMM-MRKF.
    }
\label{Fig_Crazyflie}
\end{figure}

In this part, a real-world indoor environment is used to emulate a large-scale outdoor environment.
We consider the monitoring of a non-cooperative UAV in an urban package-delivery mission, representative of emerging logistics services in AITSs.
The UAV is required to follow a predefined flight path while strictly avoiding several restricted regions, such as transportation hubs and sensitive infrastructure.
We designate $\Xi=4$ static no-fly zones, representing sensitive or hazardous areas, positioned at the following centroids:
\begin{equation}                       
\begin{aligned}
\label{eq:exp1}
    &\mathbf{p}_{1}=[0.2;1.0],
    \mathbf{p}_{2}=[0.2;-0.2], \\
    &\mathbf{p}_{3}=[1.0;-0.2],
    \mathbf{p}_{4}=[1.0;0.8].
\end{aligned}
\end{equation}
The geometry of each restricted zone is defined by its observation set $\mathcal{O}^{\mu_{\xi}}$, which are specified as:
\begin{equation}                       
\begin{aligned}
\label{eq:exp3}
    \mathcal{O}^{\mu_{1}}
    :=&
    \{\mathbf{x}\in\mathbb{R}^{6}|\|\mathbf{C}_{pos}\mathbf{x}-\mathbf{p}_{1}\|_{2}\leq 0.3)\}, \\
    \mathcal{O}^{\mu_{2}}
    :=&
    \{\mathbf{x}\in\mathbb{R}^{6}|\|\mathbf{C}_{pos}\mathbf{x}-\mathbf{p}_{2}\|_{\infty}\leq 0.2)\}, \\
    \mathcal{O}^{\mu_{3}}
    :=&
    \{\mathbf{x}\in\mathbb{R}^{6}|\|\mathbf{C}_{pos}\mathbf{x}-\mathbf{p}_{3}\|_{2}\leq 0.2)\}, \\
    \mathcal{O}^{\mu_{4}}
    :=&
    \{\mathbf{x}\in\mathbb{R}^{6}|\|\mathbf{C}_{pos}\mathbf{x}-\mathbf{p}_{4}\|_{\infty}\leq 0.2)\}.
\end{aligned}
\end{equation}
In the considered delivery scenario, we define the atomic predicate by the distance function, that is $\theta(\mathbf{x})=d_{i}-\|\mathbf{C}_{pos}\mathbf{x}-\mathbf{p}_{i}\|_{2}\ (\forall\ i)$.
Then, the constraint during the process is governed by the following STL specification:
\begin{equation}                       
\begin{aligned}
\label{eq:exp2}
    \varphi_{\xi}
    =
    \mathcal{G}_{[0,3]}(\neg\mu_{\xi})\ (\forall\ \xi=1,\cdots,4).
\end{aligned}
\end{equation}
This setup creates a challenging navigation corridor, where the non-cooperative UAV must perform precise maneuvers to satisfy the spatio-temporal constraints of the formula.
The overall experimental setup is illustrated in Fig. \ref{Fig_Crazyflie}.

We consider $M=6$ motion models in this experiment, where the angular velocities for the $4$ CT models are specified as $\omega_{CT_{1}}=-1.5\mathrm{rad/s}$, $\omega_{CT_{2}}=-0.5\mathrm{rad/s}$, $\omega_{CT_{3}}=0.5\mathrm{rad/s}$, and $\omega_{CT_{4}}=1.5\mathrm{rad/s}$.
A Vicon motion capture system is used to monitor the UAV.
Based on this, we simulate two sensors that provide only positional measurements, and an additional sensor that outputs velocity measurements, which means
\begin{equation}                       
\begin{aligned}
\label{eq:C}
    &\mathbf{C}_{1}
    =
    \mathbf{C}_{2}
    =
    \mathbf{C}_{pos}
    \triangleq
    \begin{bmatrix}
    1 & 0 & 0 & 0 & 0 & 0 \\
    0 & 0 & 0 & 1 & 0 & 0
    \end{bmatrix}, \\
    &\mathbf{C}_{3}
    =
    [\mathbf{C}_{pos};\mathbf{C}_{vel}]
    \triangleq
    \begin{bmatrix}
    1 & 0 & 0 & 0 & 0 & 0 \\
    0 & 0 & 0 & 1 & 0 & 0 \\
    0 & 1 & 0 & 0 & 0 & 0 \\
    0 & 0 & 0 & 0 & 1 & 0
    \end{bmatrix}.
\end{aligned}
\end{equation}
The frequency of monitoring is determined as $T_{\mathbf{x}}=0.1\ \mathrm{s}\ (10\ \mathrm{Hz})$, while the sampling periods of the sensors are set as $T_{\mathbf{y}_{1}}=0.2\ \mathrm{s}\ (5\ \mathrm{Hz})$, $T_{\mathbf{y}_{2}}=0.1\ \mathrm{s}\ (10\ \mathrm{Hz})$, and $T_{\mathbf{y}_{3}}=0.1\ \mathrm{s}\ (10\ \mathrm{Hz})$, respectively.
Moreover, the nominal Markov transition probability matrix is chosen as $\Pi_{k|k-1}=0.9\mathbf{I}_{M}+0.1/(M-1)(\mathbf{1}_{M}-\mathbf{I}_{M})$.
The STL robustness safety threshold is set as $\rho_{th}=0.2$.
For predictive risk evaluation, a horizon of $\delta=10\ (1\ \mathrm{s}\ )$ steps is used, and PRSs are constructed at a confidence level of $\zeta=0.05$.

\subsection{Trajectory Tracking}

\begin{figure}[htbp]
    \centering
    \subfloat[]{
        \includegraphics[width=\columnwidth]{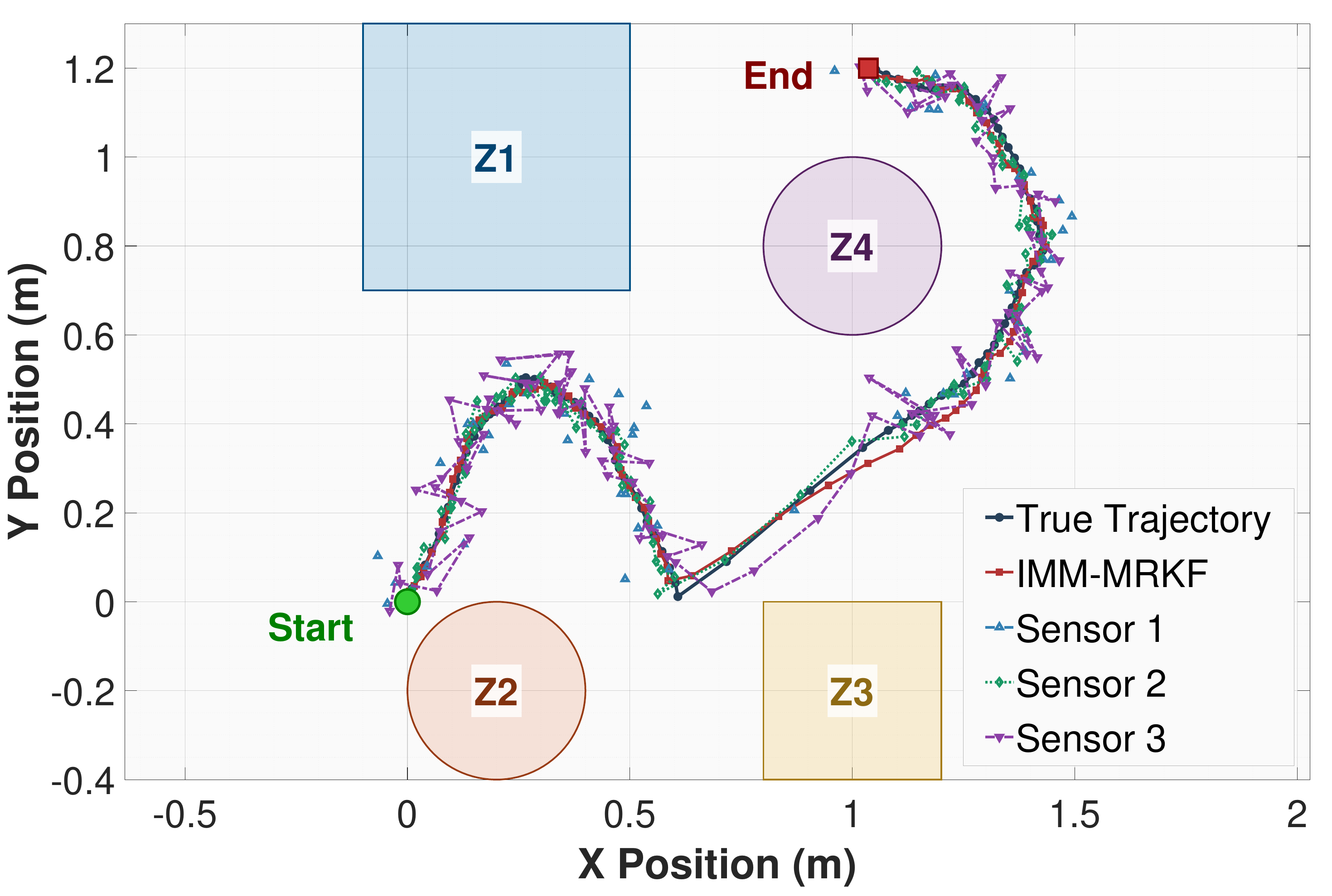}
    \label{Fig_Trajectory}
    }
    \vspace{0.1cm}
    \subfloat[]{
        \includegraphics[width=\columnwidth]{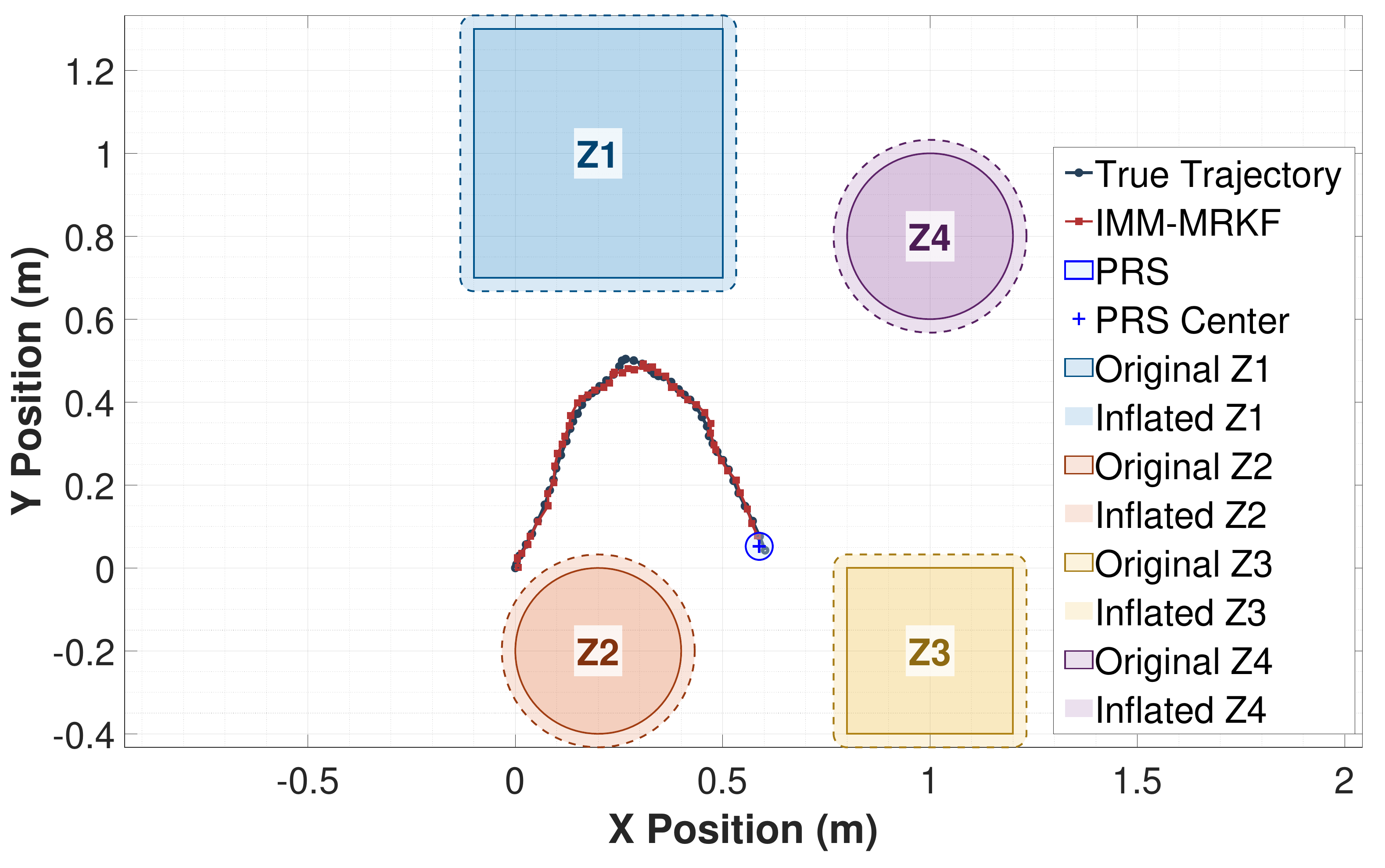}
    \label{Fig_Minkowski}
    }
    \caption{
    Experimental results of UAV trajectory estimation and PRSs.
    (a) True trajectory, fused trajectory estimated by the STL-aware IMM-MRKF, and raw measurements from the three heterogeneous sensors, shown together with the four no-fly zones.
    (b) Predicted PRSs at a certain slot, with their centers, the original no-fly zones, and the Minkowski-sum–inflated zones that represent conservative risk envelopes.
    }
\end{figure}

The trajectory tracking performance of the proposed STL-aware IMM-MRKF is evaluated against raw sensor measurements. 
Concretely, as shown in Fig. \ref{Fig_Trajectory}, the estimated trajectory (red solid line) closely aligns with the ground truth (black solid line) across all mission phases. 
Notably, this accurate tracking is maintained despite certain noise and outliers in the raw measurement data. 
These results demonstrate that fusing multi-rate measurements with multiple motion models enables a consistently reliable reconstruction of the UAV's trajectory.

To intuitively characterize the uncertainty associated with the IMM-MRKF estimate, we construct PRSs over the prediction horizon using the multi-step prediction methodology in Section IV.
Fig. \ref{Fig_Minkowski} illustrates an example of such a PRS tube, where the centers of the predicted distributions are shown together with the corresponding ellipsoidal $\zeta$-PRS (with confidence level $\zeta=0.95$).
The PRSs are visualized together with the original no-fly zones and their ``inflated'' counterparts obtained via Minkowski sum with the PRS boundary.
The inflated regions represent conservative approximations of the effective risk envelopes, i.e., spatial regions that the UAV might reach with high probability given its current state uncertainty.

\subsection{Model Probability}

\begin{figure}[htbp]
    \centering
    \subfloat[]{
        \includegraphics[width=\columnwidth]{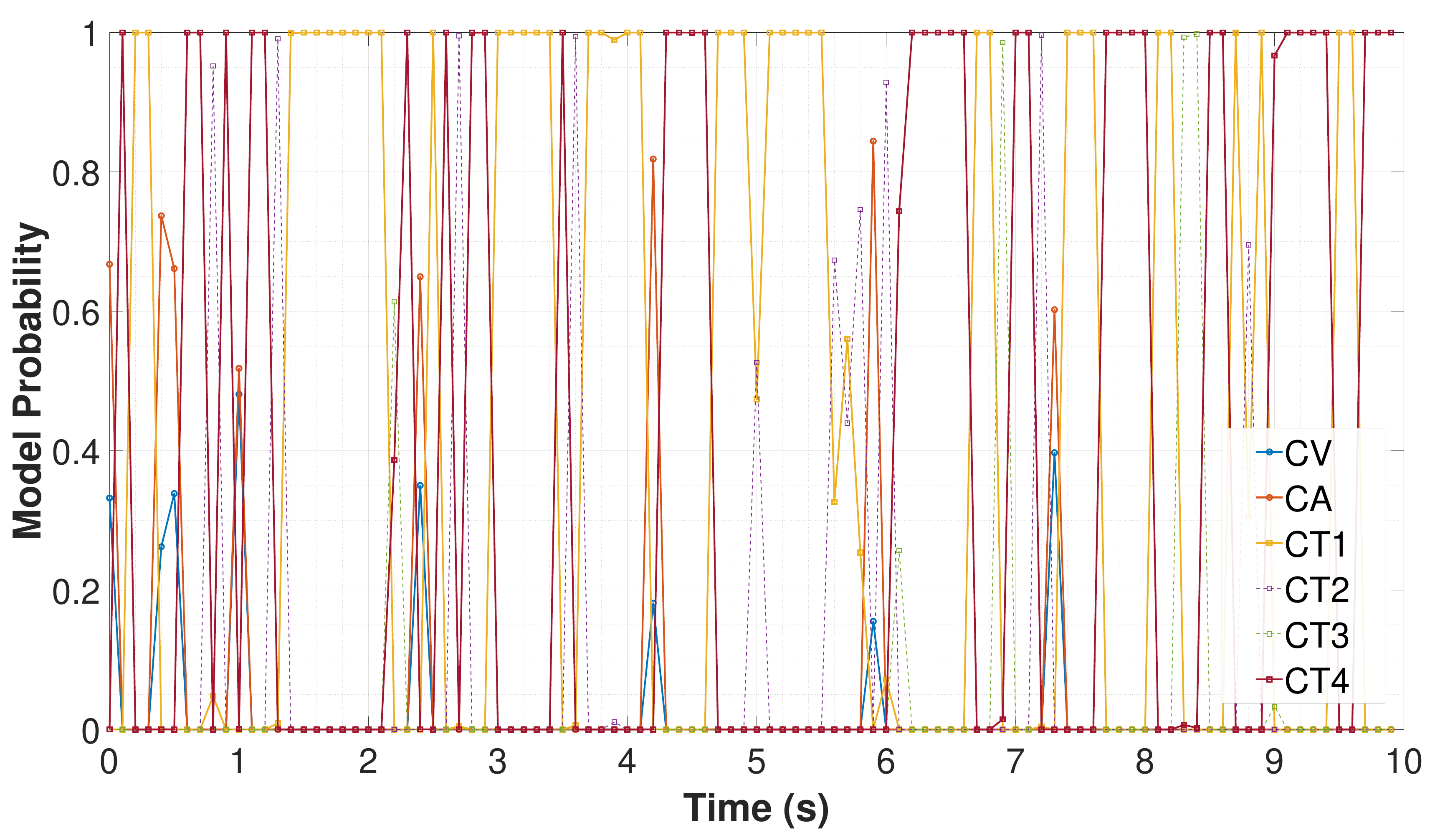}
    \label{Fig_ModelProbability}
    }
    \vspace{0.1cm}
    \subfloat[]{
        \includegraphics[width=\columnwidth]{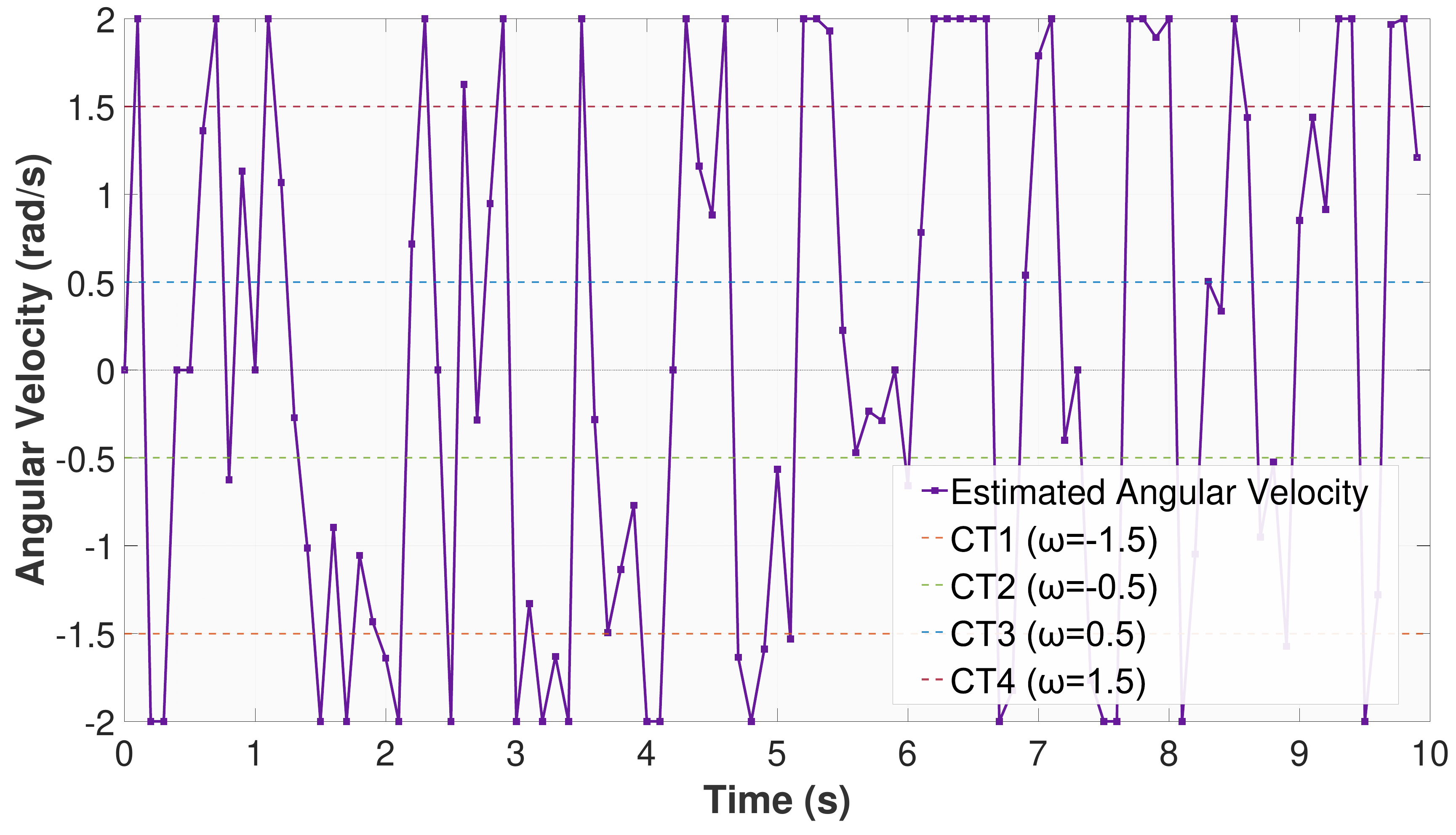}
    \label{Fig_AngularVelocity}
    }
    \vspace{0.1cm}
    \subfloat[]{
        \includegraphics[width=\columnwidth]{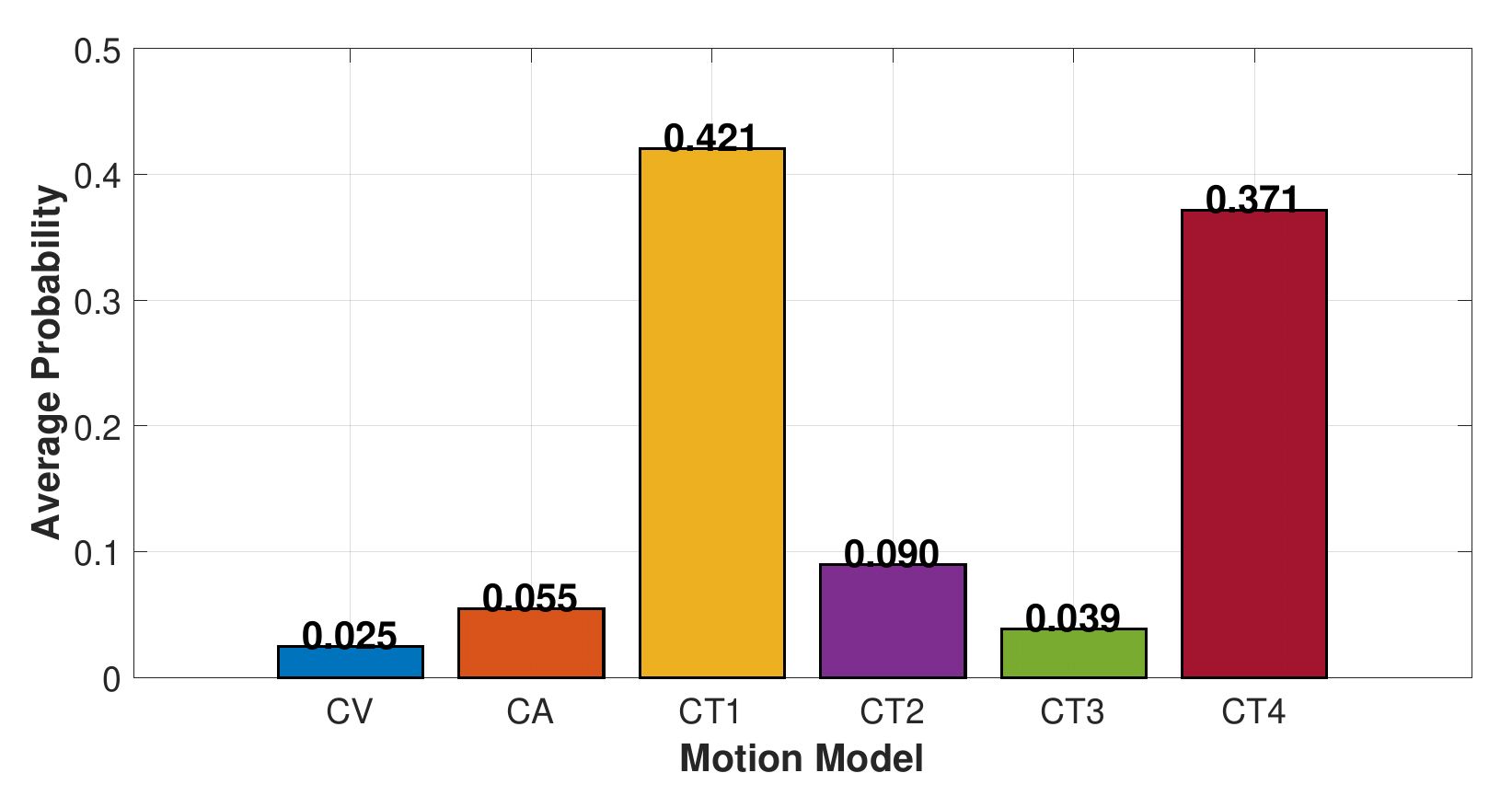}
    \label{Fig_AverageProbability}
    }
    \caption{
    Behavior of the multiple motion models (CV, CA, and four CT models) in the STL-aware IMM-MRKF.
    (a) Temporal evolution of model probabilities, illustrating automatic adaptation to turning and straight-flight phases.
    (b) Estimated angular velocity over time, correlated with activation of CT models.
    (c) Average probability of each model across the entire experiment, summarizing the dominant motion patterns.
    }
\end{figure}

Notice that a key feature of the proposed framework is its STL-aware, kinematics-informed model transition mechanism, which dynamically adjusts the probabilities of the candidate motion models based on both estimated motion characteristics and STL robustness.
Hence, we examine how model probabilities evolve over time and how they correlate with the UAV’s maneuvers and proximity to restricted regions.

Fig. \ref{Fig_ModelProbability} depicts the temporal evolution of the model probabilities in the proposed STL-aware IMM-MRKF framework. 
The adaptation process can be roughly categorized into three distinct operational phases based on the motion pattern.
During the initial phase, the UAV executes a clockwise circular trajectory.
The model probability for the constant-turn model CT1 remains consistently high, confirming the identification of the UAV's turning maneuver.
Then, the UAV transitions to a near-straight path with sustained acceleration.
Consequently, the probability of the CA model rises, reflecting the filter's adaptive response to the change in kinematic mode.
In the final phase, the UAV performs a counterclockwise circular turn, inverse to that in the initial phase. 
The filter accordingly shifts its weight to the CT4 model, whose probability peaks during this interval.
It also demonstrates the capability to track reversing turning behaviors.



Fig. \ref{Fig_AngularVelocity} shows the temporal evolution of the estimated angular velocity along with the corresponding model probabilities.
A clear correlation is evident, where peaks in the estimated angular velocity correspond to a significant increase in the probability of the corresponding CT models. 
Conversely, periods of near-zero angular velocity are primarily associated with the CA model, with occasional contributions from the CV model during straight-line flight.
This alignment demonstrates that the proposed STL-aware transition mechanism effectively translates low-level kinematic cues into high-level model belief updates.

Fig. \ref{Fig_AverageProbability} depicts the overall proportion of each motion model across the entire estimation horizon.
This distribution reveals which models contributed most significantly to the fused state estimate.
For example, the predominance of the CA model would indicate extended straight-flight segments, while substantial shares of CT models correspond to periods of active maneuvering.
This aggregated statistical profile complements the time-varying probabilities in Fig. \ref{Fig_ModelProbability}, and it concisely summarizes the UAV's dominant operational modes and the filter's adaptive model engagement.

In summary, the adaptation of model probabilities demonstrates that the proposed STL-aware IMM-MRKF framework can actively integrate high-level safety semantics with low-level kinematics.
It will guide the model selection in a risk-sensitive manner rather than merely reacting to local motion.
The closed-loop interaction between STL robustness and model probabilities lies at the core of the proposed framework's capability to maintain reliable estimation, particularly within safety-critical regions near operational boundaries.


\subsection{STL Robustness, Risk, and Warning Level}

\begin{figure}[htbp]         
\centering
    \includegraphics[width=\columnwidth]{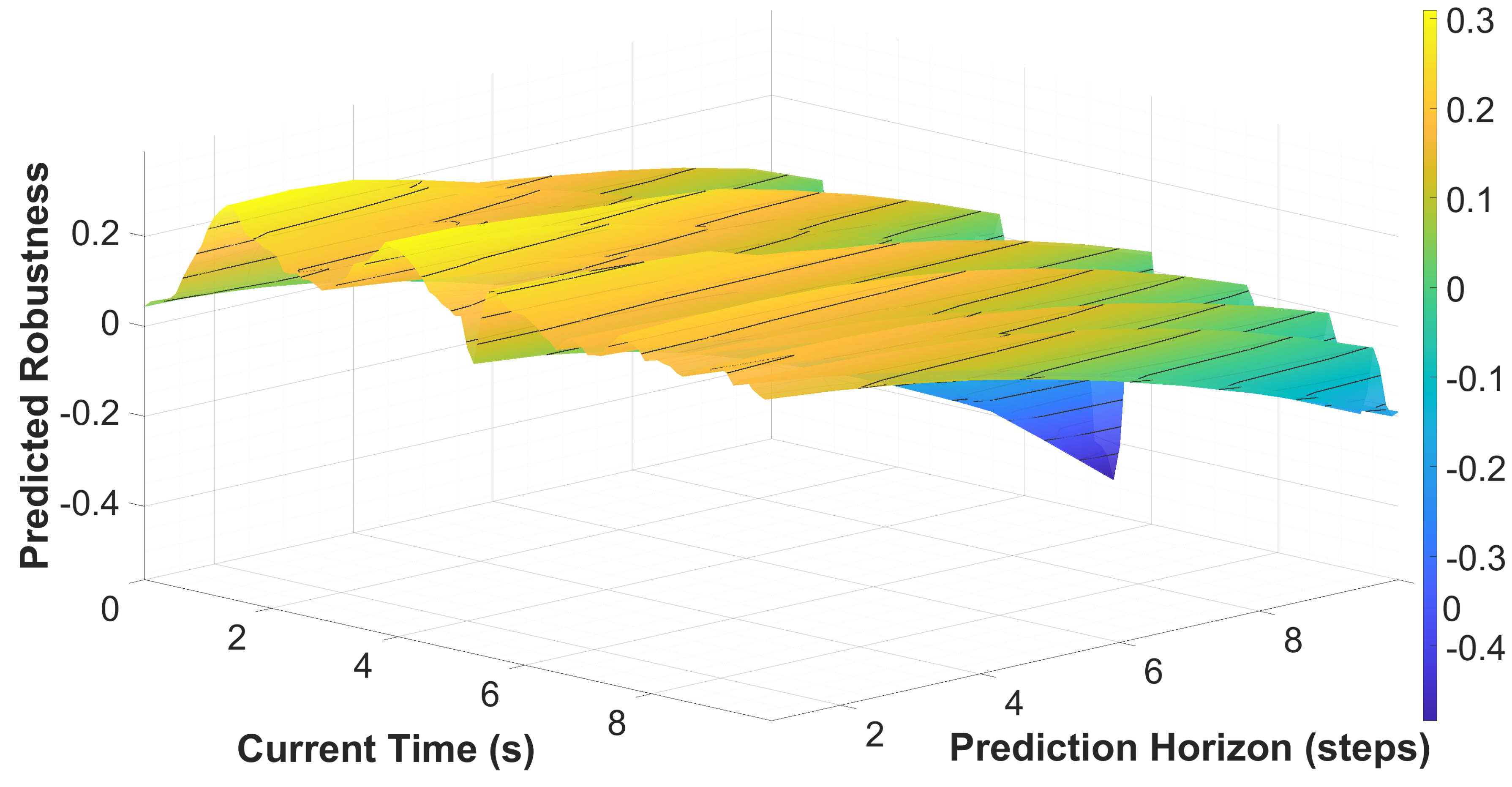}
    \caption{
    Predicted STL robustness over the current time and the prediction horizon. 
    Darker regions indicate lower robustness (higher anticipated risk of violating no-fly-zone specifications), showing that the framework can foresee future unsafe conditions before they occur.
    }
\label{Fig_Prediction}
\end{figure}

\begin{figure}[htbp]
    \centering
    \subfloat[]{
        \includegraphics[width=\columnwidth]{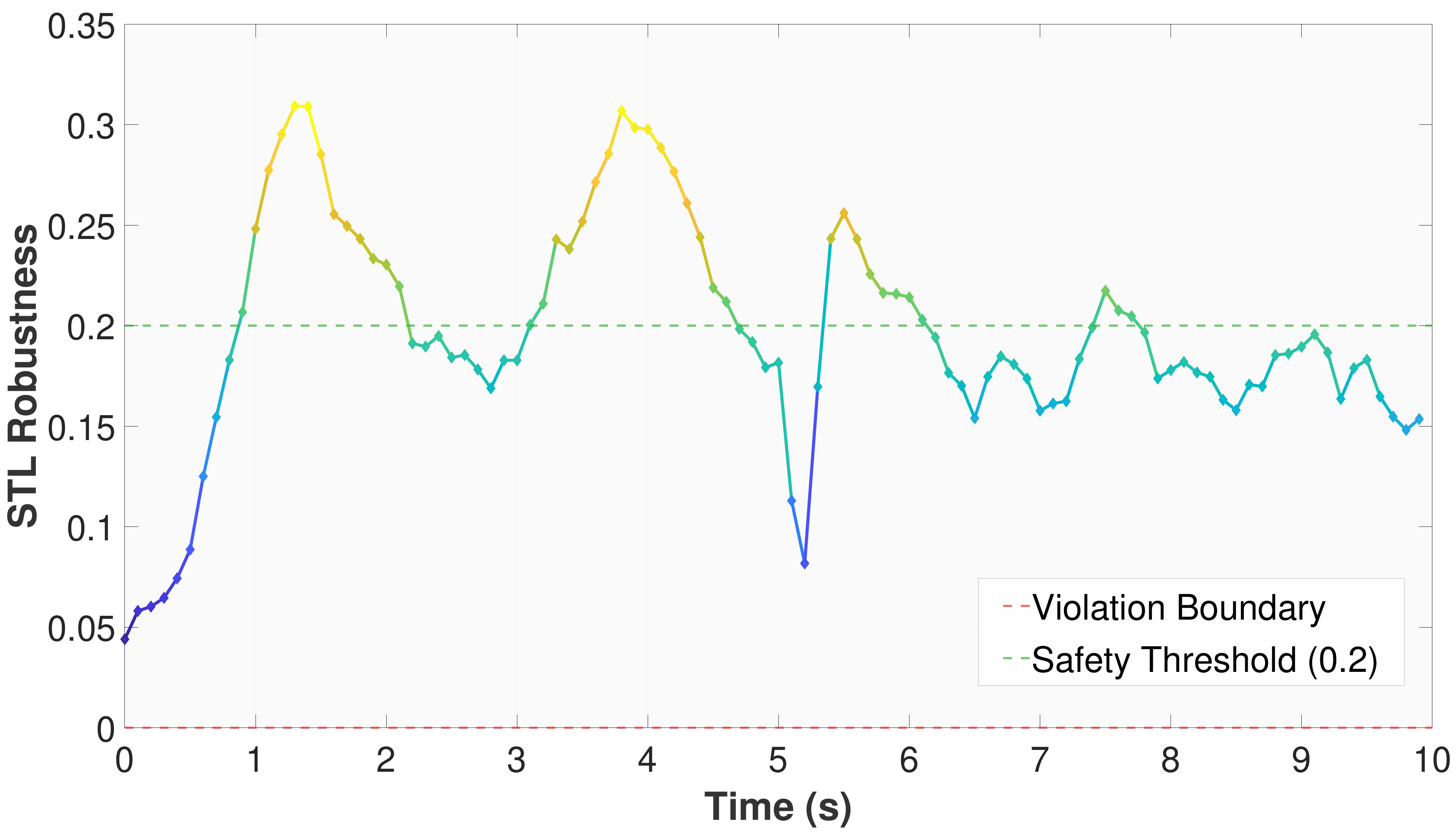}
    \label{Fig_Robustness}
    }
    \vspace{0.1cm}
    \subfloat[]{
        \includegraphics[width=\columnwidth]{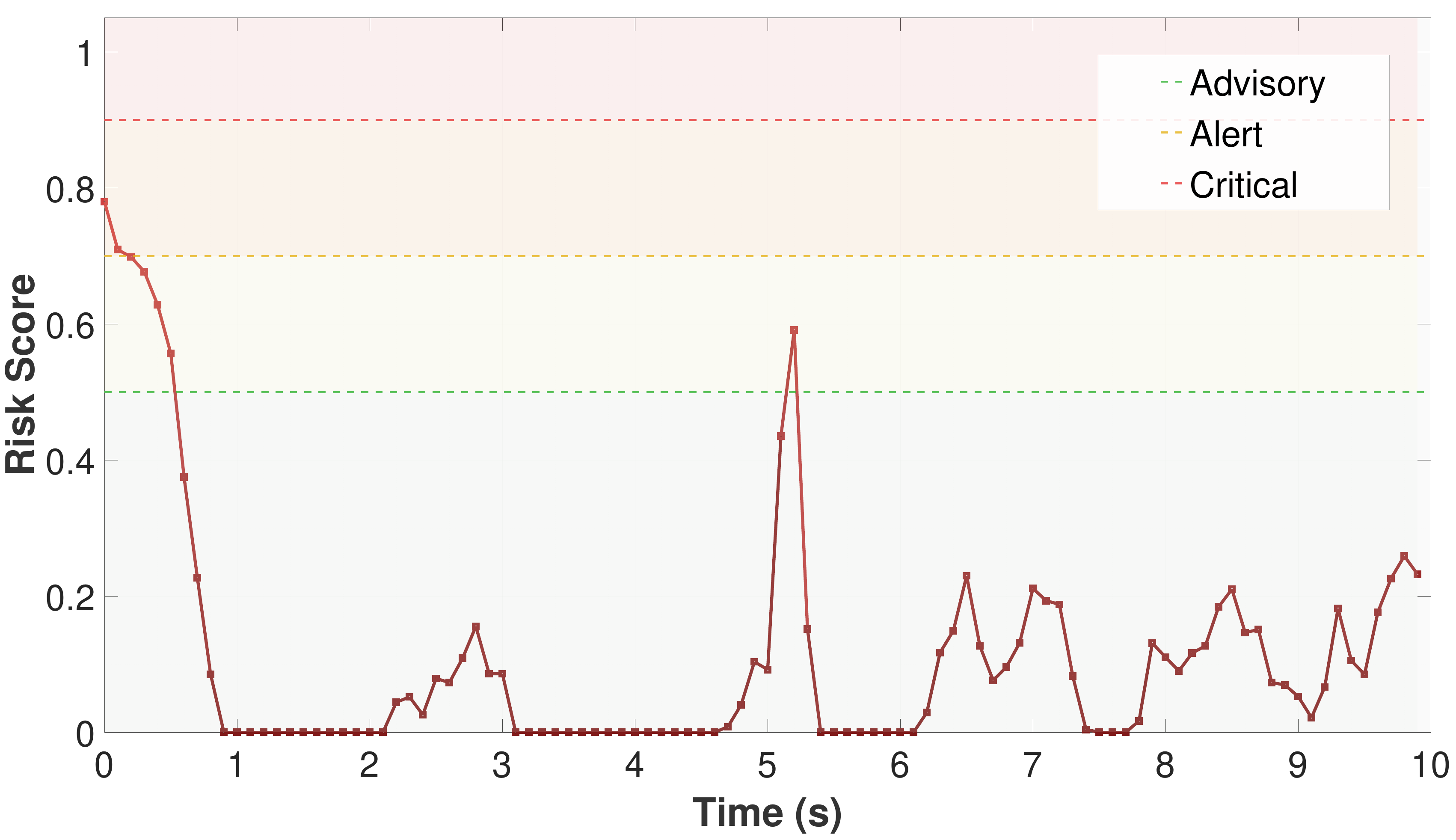}
    \label{Fig_Risk}
    }
    \vspace{0.1cm}
    \subfloat[]{
        \includegraphics[width=\columnwidth]{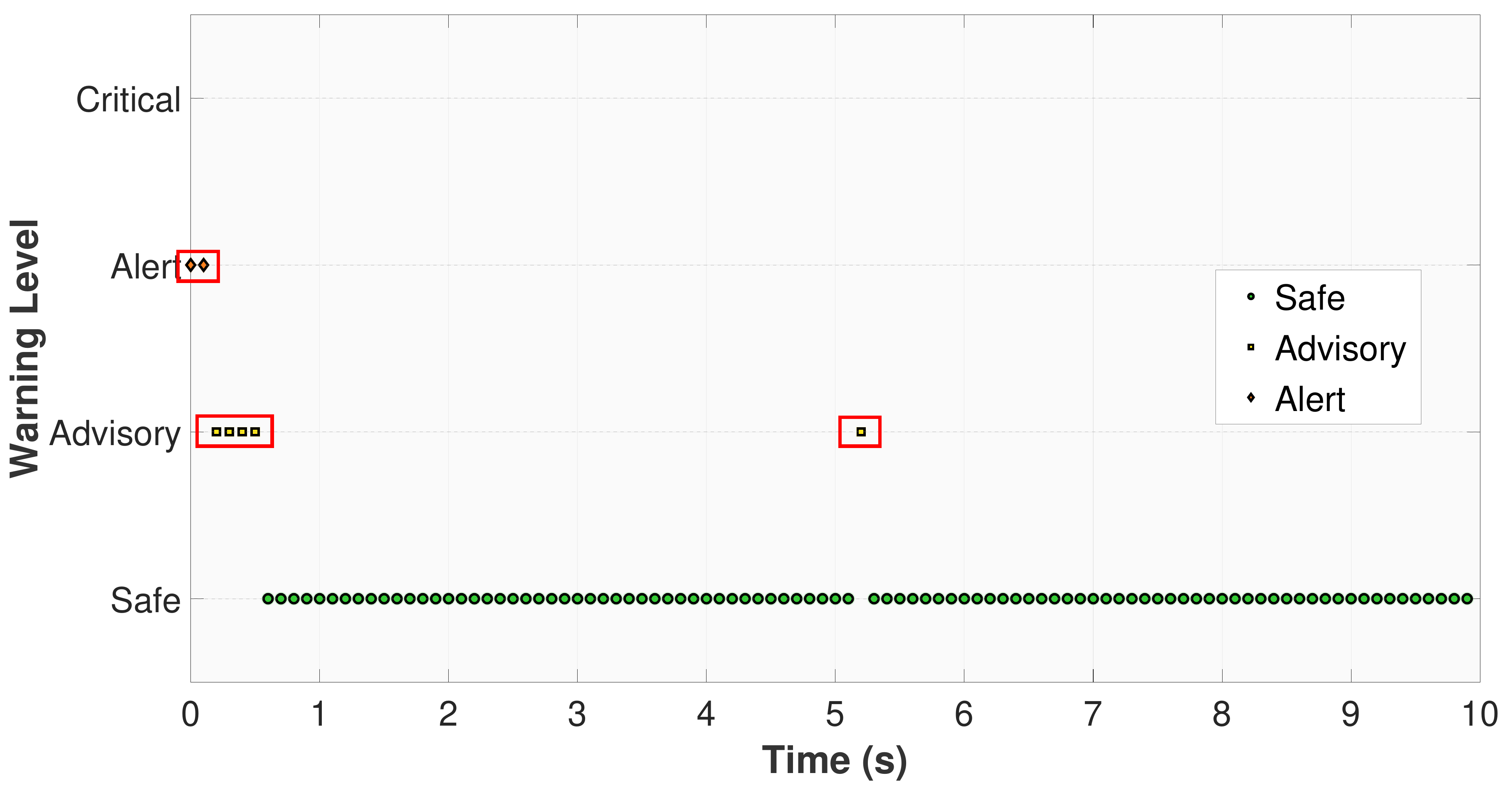}
    \label{Fig_Warning}
    }
    \caption{
    STL robustness, risk score, and warning levels for the UAV delivery experiment.
    (a) Real-time robustness with respect to the no-fly-zone specifications, together with the violation boundary and safety threshold.
    (b) Derived risk score and the thresholds defining Advisory, Alert, and Critical levels.
    (c) Discrete warning level over time, showing automatic escalation from Safe to higher levels as predicted risk increases.
    }
\end{figure}

Then, we examine how the integration of STL-based robustness semantics, probabilistic reachability, and a multi-tiered warning logic enables predictive safety assessment and timely alert generation. 
Following the robustness semantics defined in Section IV, the robustness of both the estimated and predicted trajectories is evaluated against the STL specification of each no-fly zone. 
At each time step, the state distribution is propagated over a finite prediction horizon of $\delta=10$ steps to construct the corresponding PRS tubes. 
For every predicted time $\tau\in(k,k+\delta]$, robustness is then evaluated over the reachable set, producing a forward-looking robustness profile.

Fig. \ref{Fig_Prediction} visualizes the predicted robustness across the current time and the prediction horizon. 
In this representation, darker regions correspond to lower robustness (higher risk), while lighter regions indicate safer configurations with larger margins from the no-fly zones.
Decreases in robustness are observed before the UAV reaches the vicinity of restricted regions in the nominal trajectory, because of the PRS inflation capturing the potential spread of future states.
As a result, the framework anticipates possible violation to no-fly zones several steps ahead in time, rather than only at the current time instant.
Nonetheless, there are intervals where instantaneous robustness remains comfortably positive, while predicted robustness over the horizon deteriorates. 
This gap highlights the value of predictive analysis: even when the current state is safe, future uncertainty and motion dynamics may lead to elevated risk.

Next, we examine the derived risk metric and the resulting warning levels. 
Fig. \ref{Fig_Robustness} depicts the evolution of real-time robustness, along with the violation boundary and the safety threshold.
The robustness declines as the UAV enters the constrained corridor near the no-fly zones, while recovers upon exiting this high-risk region.
Instances where robustness approaches the threshold correspond to configurations where the boundaries of the inflated no-fly zones and the PRSs become nearly tangent, which indicates the heightened risk of constraint violation.

Fig. \ref{Fig_Risk} plots the time evolution of the risk score with horizontal lines marking the thresholds that define the Advisory, Alert, and Critical warning levels.
As anticipated, the risk score remains close to $0$ when the UAV operates far from restricted areas, increases as the STL robustness degrades, and peaks when the PRS exhibits significant overlap with the inflated no-fly zones.
Risk increases to the range $[0.5, 0.7)$ correspond to plausible future violations, prompting advisory-level alerts.
Peaks within $[0.7, 0.9)$ signal a high likelihood of violation within the prediction horizon, where accumulated motion uncertainty elevates the risk.

Fig. \ref{Fig_Warning} shows the resulting discrete warning level over time.
As the robustness decays and the UAV nears the inflated zone boundaries, the system will transition to the Advisory state. 
This shift occurs when the UAV remains at a safe distance from the actual no-fly zones, which provides operators and high-level planners with sufficient time to consider corrective actions or enhanced monitoring.
When the PRS intersects more substantially with the inflated zones, representing a heightened probability of violation, the warning level escalates to Alert and, in critical cases, to Critical.
These elevated levels correspond to scenarios in which an immediate response, such as trajectory replanning or active countermeasures, is warranted.


Overall, the results demonstrate that the proposed framework successfully transforms multi-model uncertainty‑aware state estimates into actionable, tiered risk evaluations. 
By integrating STL robustness, probabilistic reachability, and a structured warning logic, the system can provide reliable predictive alerts ahead of potential airspace violations, which significantly improves situational awareness and facilitates proactive safety management.





\subsection{Performance Analysis}


\begin{figure}[htbp]
    \centering
    \subfloat[]{
        \includegraphics[width=\columnwidth]{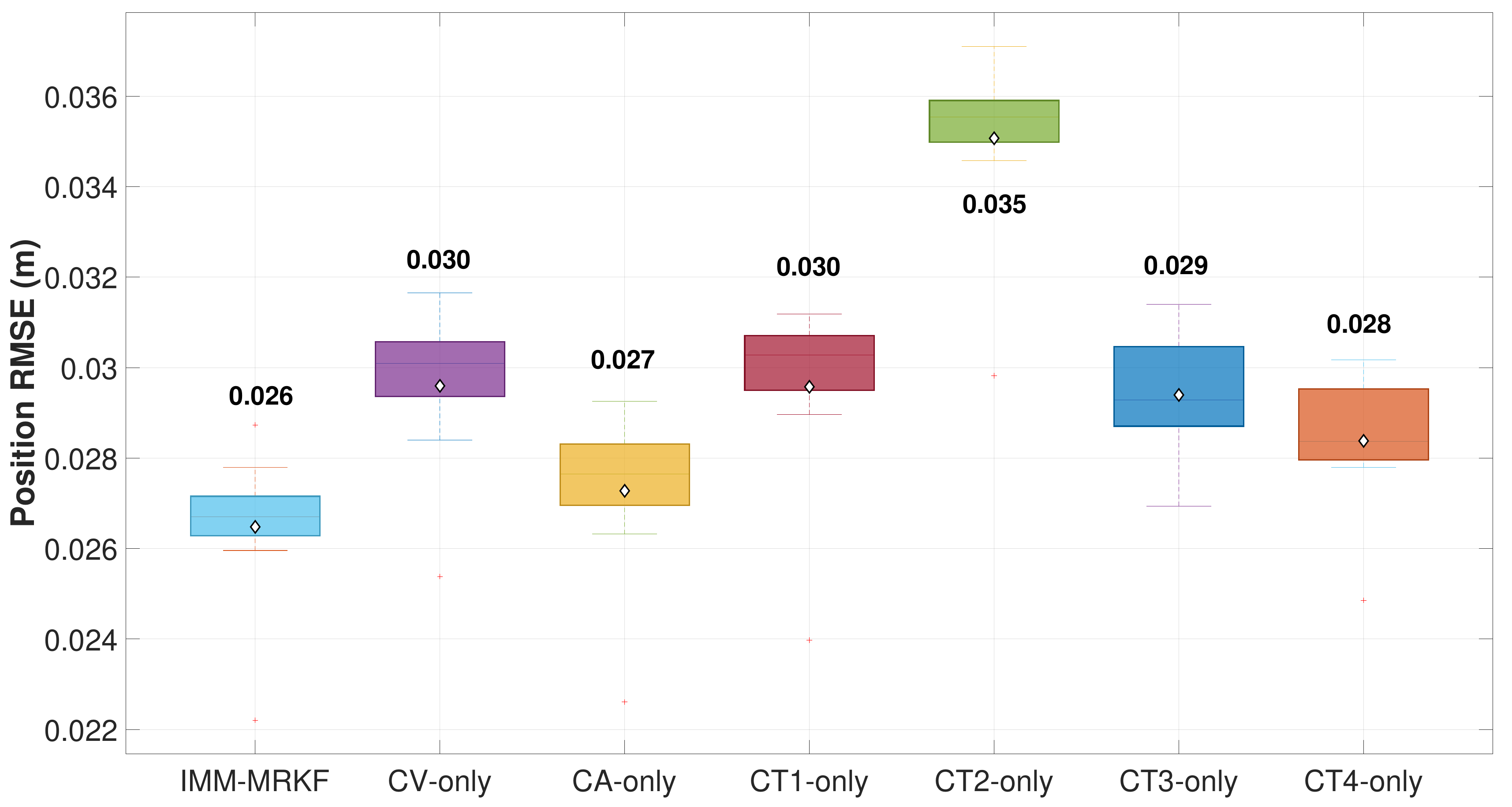}
    \label{Fig_RMSE}
    }
    \vspace{0.1cm}
    \subfloat[]{
        \includegraphics[width=\columnwidth]{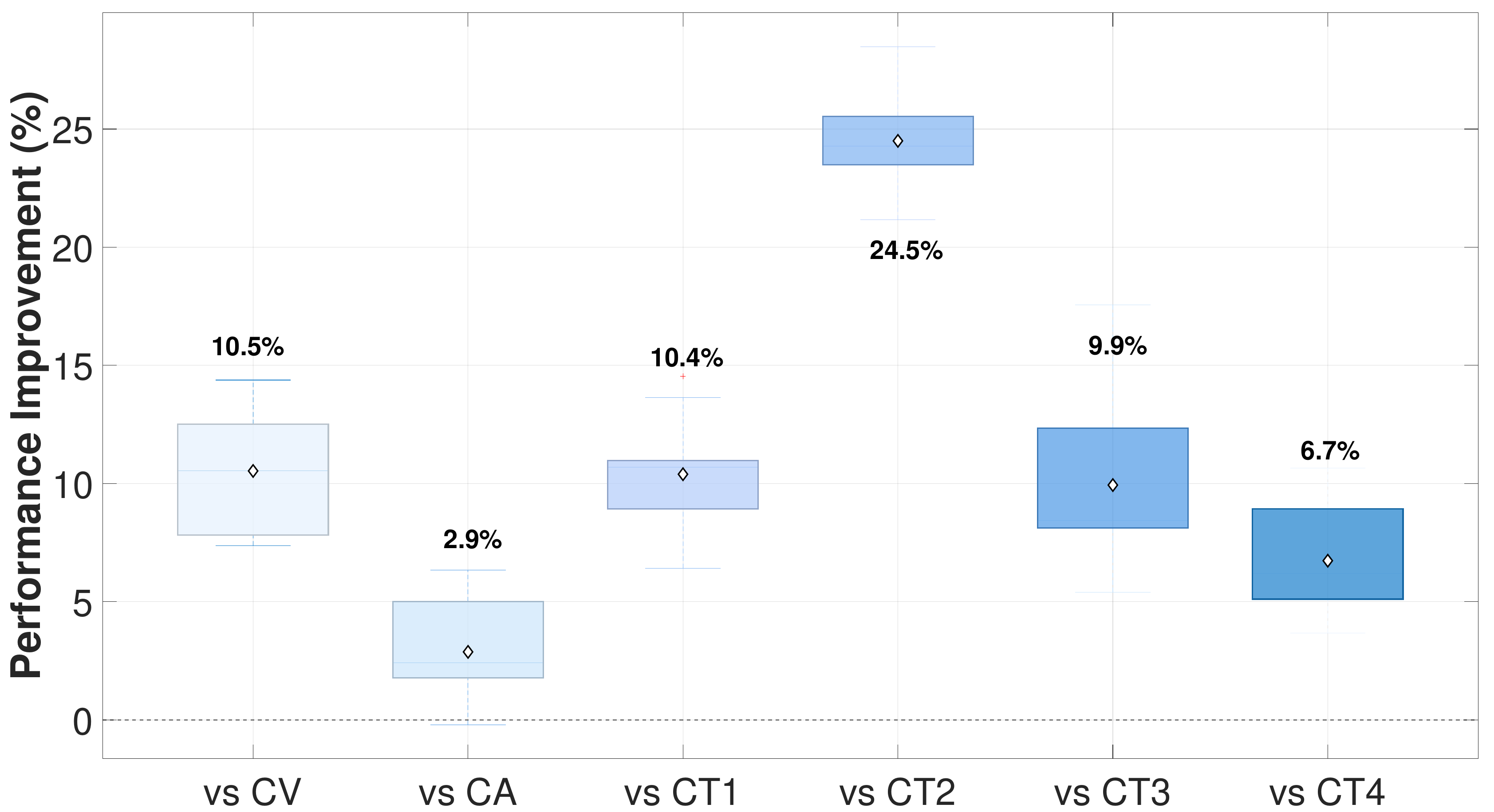}
    \label{Fig_RMSE_Improvement}
    }
    \caption{
    Monte Carlo comparison of the proposed STL-aware IMM-MRKF and single-model Kalman filters.
    (a) Position RMSE boxplots for IMM-MRKF and CV-/CA-/CT-only filters, where IMM-MRKF achieves the lowest error and smallest dispersion.
    (b) Percentage RMSE improvement of IMM-MRKF over each single-model filter, demonstrating consistent performance gains across runs.
    }
\end{figure}

To provide a comprehensive statistical evaluation of the proposed framework's performance and robustness, a Monte Carlo analysis consisting of 1000 independent runs is conducted.
The analysis focuses on comparing the root mean square error (RMSE) distributions of the proposed STL‑aware IMM‑MRKF against those of several single‑model Kalman filters (CV, CA, CT1, CT2, CT3, CT4).

Fig. \ref{Fig_RMSE} shows a boxplot comparison of the position RMSE for the seven filters.
Each box summarizes the RMSE distribution obtained from the Monte Carlo runs for a given filter, with the lower and upper edges representing the first ($Q1$) and third ($Q3$) quartiles, respectively.
The central line marking the median, and the whiskers extending to the minimum and maximum non‑outlier values, while the diamond symbol denotes the arithmetic mean RMSE.

The proposed IMM‑MRKF achieves the lowest median RMSE ($0.026$) and the smallest interquartile range ($IQR=Q3-Q1=0.002$), indicating that it consistently delivers the most accurate and stable state estimates across all runs. 
In contrast, the RMSE distributions of the single‑model filters are substantially more dispersed, especially for the CV and CT models, as evidenced by their wider boxes and the presence of multiple upper outliers. 
This dispersion underscores the sensitivity of single model filters to motion mode mismatch. 

To quantify the performance gain of the IMM-MRKF over each single model, we computed the percentage reduction in RMSE for each simulation run:
\begin{equation}                       
\begin{aligned}
\label{eq:exp4}
    \mathrm{Improvement}_{m}
    =
    \frac{\mathrm{RMSE}_{m}-\mathrm{RMSE}_{\mathrm{IMM}}}{\mathrm{RMSE}_{m}}\times 100\%\ (\forall\ m).
\end{aligned}
\end{equation}
Fig. \ref{Fig_RMSE_Improvement} displays the distribution of these performance improvements in a boxplot format. 
Note that the entire interquartile range for every model comparison lies above the zero‑improvement line, and the 25th percentile for each exceeds $5\%$.
This pattern provides strong evidence that the performance gain of the IMM‑MRKF over every single‑model filter is significant at the $95\%$ confidence level.

In summary, the systematic Monte Carlo evaluation statistically proves the core contributions of the proposed framework.
Through the integration of STL semantics with multiple‑model estimation, we achieve accurate and safety‑aware tracking in uncertain settings and deliver reliable performance for predictive risk warning.

\section{Conclusion} 

This paper has presented a unified online monitoring and predictive warning framework for non-cooperative UAVs by integrating an IMM-MRKF with STL. 
The proposed method addresses two key challenges: accurate state estimation under uncertain, time-varying motion modes, and predictive risk assessment with respect to formally specified spatio-temporal constraints. 
An STL-aware model transition mechanism embeds safety semantics directly into the filter’s adaptation logic, while PRSs provide a forward-looking quantification of violation risk.
Experimental validation in a UAV delivery scenario has shown that the framework achieves robust multi-sensor tracking, context-sensitive model adaptation, and timely multi-level warnings.
Future work will focus on extending the framework to three-dimensional motion, richer sensing modalities, and larger-scale constrained multi-agent scenarios.
Another important direction is the analysis of stability, conservatism, and computational tradeoffs induced by robustness-informed mode adaptation.


\vfill

\end{document}